\documentclass[a4paper]{article}

\usepackage[english]{babel}
\usepackage{a4wide}
\usepackage[utf8x]{inputenc}
\usepackage{amsmath,amssymb,marvosym}
\usepackage{graphicx,color}
\usepackage{caption,subcaption}

\usepackage{tikz}
\usetikzlibrary{calc,decorations.markings}
\usetikzlibrary{snakes,arrows}
\usetikzlibrary{decorations.pathmorphing, patterns,shapes}

\usepackage{hyperref}

\newcommand{\eq}[1]{\begin{equation}#1\end{equation}}

\newcommand{\lt}{l}
\newcommand{\rt}{r}

\newcommand{\twist}[1][]{\mathcal{T}_{#1}}
\newcommand{\atwist}[1][]{\tilde{\mathcal{T}}_{#1}}
\newcommand{\logneg}{\mathcal{E}}
\newcommand{\chtwist}[1][]{\tau_{#1}}
\newcommand{\chatwist}[1][]{\tilde{\tau}_{#1}}

\newcommand{\cdim}[1][n]{\Delta_{#1}}
\newcommand{\cdimsq}[1][n]{\Delta^{(2)}_{#1}}

\newcommand{\Ctttsq}{C_{\twist\twist}^{\twist^2}}
\newcommand{\Ctatsqat}{C_{\twist\atwist^2}^{\atwist}}
\newcommand{\Ctatsqt}{C_{\twist\atwist^2\twist}}

\definecolor{light blue}{RGB}{170,200,255}
\definecolor{lilac}{RGB}{200,100,255}
\definecolor{light red}{RGB}{255,80,80}

\begin{document}

\begin{center}

{\Large {\bf Entanglement negativity and entropy in non-equilibrium conformal field theory}}

\vspace{0.8cm} 

{\large Marianne Hoogeveen${}^{\spadesuit}$ and \text{Benjamin Doyon${}^{\spadesuit}$}}

\vspace{0.2cm}
{\small ${}^{\spadesuit}$ Department of Mathematics, King's College London,
London, United Kingdom.}
\end{center}

\vspace{0.5cm} 
We study the dynamics of the entanglement in one dimensional critical quantum systems after a local quench in which two independently thermalized semi-infinite halves are joined to form a homogeneous infinite system and left to evolve unitarily. 
We show that under certain conditions a nonequilibrium steady state (NESS) is reached instantaneously as soon as the entanglement interval is within the light-cone emanating from the contact point. In this steady state, the exact expressions for the entanglement entropy and the logarithmic negativity are in agreement with the steady state density matrix being a boosted thermal state, as expected.
We derive various general identities: relating the negativity after the quench with unequal left and right initial temperatures, with that with equal left and right temperatures; and relating these with the negativity in equilibrium thermal states. In certain regimes the resulting expressions can be analytically evaluated. Immediately after the interval interesects the lightcone, we find logarithmic growth. For a very long interval, we find that the negativity approaches a plateau after sufficiently long times, different from its NESS value. This provides a theoretical framework explaining recently obtained numerical results.

\medskip

\hfill \today
\medskip


\section{Introduction}
Finding ways to quantify the entanglement of quantum many body systems is an interesting problem with various applications, for instance as a tool for detecting quantum critical behaviour, and topolocial phases  \cite{AmicoOsterlohVedral2008review, eisert2010area}.
A measure of the quantum entanglement for bipartite systems in a pure state is the Entanglement Entropy (EE),
\eq{\label{eq:EE}
S_A=-\text{Tr}\rho_A\ln\rho_A,
}
which is calculated using the reduced density matrix $\rho_A=\text{Tr}_B\rho$, where $\rho$ is the density matrix of the whole system, and $A$ and $B$ are complementary parts of the system. 
Together with the R\'{e}nyi entropies,
\eq{\label{eq:Renyi}
S_A^{(n)}=\frac{1}{1-n}\ln\text{Tr}\rho_A^{n},\qquad \lim_{n\rightarrow 1}S_A^{(n)}=S_A,
}
this 
encodes a lot of information about the entanglement \cite{JonathanPlenio1999,Nielsen1999,Turgut2007}.
A striking feature of the EE is the universal behaviour it displays near a Quantum Phase Transition. This allows one to compute it using methods from Quantum Field Theory, or, exactly at criticality, Conformal Field Theory. In \cite{HolzheyLarsenWilczek1994,CalabreseCardy2004}, a field theory method was introduced to compute the entanglement entropy using the replica trick,
whereby $\text{Tr}\rho_A^n$ is interpreted as a partition function on an $n$-sheeted Riemann surface.

%
%
When a system is in a mixed state, the EE is not a good measure of entanglement, as it contains a classical contribution from the entropy of the mixed state. A measure of entanglement that does not have this problem for mixed states is the \textit{logarithmic negativity} \cite{VidalWerner2002}. The logarithmic negativity between two parts $A_1$ and $A_2$ (such that $A=A_1\cup A_2$, and the total system is $A_1\cup A_2\cup B$) is given by
\eq{
\logneg_{A_1,A_2}\equiv \ln||\rho_A^{T_2}||_1=\ln\text{Tr}|\rho_A^{T_2}|,
}
where the trace norm $||\rho_A^{T_2}||_1$ is the sum of the absolute values of the eigenvalues $\lambda_i$ of $\rho_A^{T_2}$, and $\rho_A^{T_2}$ is the partial transpose of $\rho_A$ with respect to the tensor factor corresponding to $A_2$. Note that $A_1$ and $A_2$ need not be complementary parts of the system (which is the case only when $B=\emptyset$).
In \cite{CalabreseCardyTonni2012, CalabreseCardyTonni2013}, a systematic way was developed to compute the logarithmic negativity using field theory methods. 
%
%
%
%
%

Recently, the dynamics of entanglement out of equilibrium has seen a surge in interest, and
this has been studied in a variety of cases. A system can be brought out of equilibrium by applying a quantum quench, which is a sudden change of a parameter in the Hamiltonian, such that the new Hamiltonian does not commute with the original one. This can be a global quench, such as a sudden change of a mass parameter, external magnetic field or interaction strength
, or a local quench, such as a sudden change in interaction strength between two sites on a chain. Such situations offer insight into quantum physics out of equilibrium.
A particular type of local quench is the so-called ``cut and glue'' quench, in which a system is cut into two pieces and glued together, possibly after the separate halves have been thermalized at different temperatures. This type of quench has been studied since a long time, especially from the viewpoint of constructing non-equilibrium steady states (where the setup is referred to as the ``partitioning approach'')  \cite{CaroliCombescotNozieresSaint-James1971,RubinGreer1971,SpohnLebowitz1977}.
Numerical results for the EE after this type of quench, in which two infinite chains of free fermions in their respective ground states are connected at a point, were found in \cite{EislerPeschel2007}.
Analytical results for the more general case where the theories could be described by a CFT  were found in \cite{CalabreseCardy2007} (with a correction in \cite{CalabreseCardy2009}). 
%
These results were generalized for systems of finite length \cite{StephanDubail2011}, and re-obtained by developing the holographic dual of the local quench between two CFTs in their ground state \cite{FarajiEsmaeil2014}.
In the same setup, the mutual information and EE after a local quench at zero temperature were calculated in \cite{AsplundBernamonti2013}, and in a nonequilibrium steady state in the presence of an energy current (in an infinite chain of free fermions) \cite{EislerZimboras2014}. An analytical formula for the negativity after the quench was conjectured for the case in which the system reaches a nonequilibrium steady state with finite energy current \cite{EislerZimboras2014b}, and numerical results were found for any time after the quench, confirming the relation for the NESS.

In this paper we consider the entanglement arising after the ``cut and glue'' quench in CFT with independently thermalized halves, where an energy current is generated and a nonequilibrium steady state is reached at late times \cite{bernard2012energy}. We confirm and generalize to a certain class of CFTs the results of \cite{EislerZimboras2014,EislerZimboras2014b}. We find that in certain CFTs, at any time, the negativity after the quench at different temperatures can be written in terms of negativities in systems where the temperatures are equal. We also find equations for various time regimes relating the negativity after the quench with equilibrium expressions, in which the effect of the time evolution is only present in a change of the intervals. For certain time regimes, we find analytical results for the finite time behaviour before the NESS is reached.

We present new techniques using holomorphic (chiral) twist fields for computing measures of entanglement after a local quench, expanding previous results  \cite{CalabreseCardy2004,CardyCastro-AlvaredoDoyon2007} (see also the various reviews in \cite{CalabreseCardy2009}). 
We consider both R\'{e}nyi entropies and the logarithmic negativity, but we concentrate on the latter, as the former is not a good measure of entanglement when considering mixed states.

The organisation of the paper is as follows: in section~\ref{sec:localquench}, we describe the type of ``cut and glue" quench we are interested in, and give an overview of important results in the literature, as well as the main results obtained in this paper. In section~\ref{sec:twistfields}, we review the properties of branch-point twist fields, and discuss \textit{holomorphic twist fields}. We use these fields to find a relation between the EE after the quench in terms of equilibrium quantities. In section~\ref{sec:evolution} we finally describe the time evolution of the logartihmic negativity after a local quench using these holomorphic twist fields, and identify cases in which a universal result can be obtained. In particular, we find results for the nonequilibrium steady state (NESS), and the time evolution leading up to that. In Appendix~\ref{sec:scattering} we confirm that the NESS, on (holomorphic) twist fields, can still be described \cite{bernard2012energy,BhaseenDoyonLucasSchalm2013} as a boosted thermal state. In Appendix~\ref{sec:structureconstants} we find relations between structure constants that appear in OPEs of twist fields, in Appendix~\ref{sec:boundary_entropy} we relate some of the nonuniversal constants in our results to the boundary entropy, and in Appendix~\ref{sec:mutualinfo} we relate the time evolution of the mutual information to quantities that can be computed in equilibrium.


\section{Local quench between independently thermalized systems}
\label{sec:localquench}

Consider two copies of a semi-infinite one-dimensional quantum system (say, a spin chain), separately prepared in generically different thermal states with inverse temperatures $\beta_l$ and $\beta_r$. As the two copies are separately prepared, they are  completely unentangled. At time $t=0$, the two copies are connected at their boundary point, forming the left and right halves of a single infinite, homogeneous total system. The total system is then left to evolve unitarily. The process we are describing is 
a local quench: at time $t=0$, the dynamics is suddenly changed from that of two disconnected semi-infinite systems to that of a single homogeneous total system, by adding a local connection (one or a few links in the spin chain). Because of the interaction thus created between the left and right subsystems, one expects the subsystems to become entangled as time goes on.

In the total system, energy can flow between the left and right subsystems, and because of the initial temperature imbalance an energy current develops. In certain situations, including those we will be considering here, after an infinitely long time a steady state emerges where energy flows constantly from one side to the other \cite{Tasaki2000,Tasaki2001,HoAraki2000,Ogata2002,AschbacherPillet2003,bernard2012energy,ColluraKarevski2014,ColluraMartelloni2014,DoyonLucasSchalmBhaseen2014,DeLucaMartelloniViti2014,Doyon2014}. This quench process is sometimes referred to as the ``partitioning approach'' for constructing non-equilibrium steady states (NESS) \cite{CaroliCombescotNozieresSaint-James1971,RubinGreer1971,SpohnLebowitz1977}. We are interested in the dynamics of the entanglement between the left and right subsystems in the presence of a developing energy current, and in the entanglement present in the steady state emerging at late times.


\begin{figure}[here]
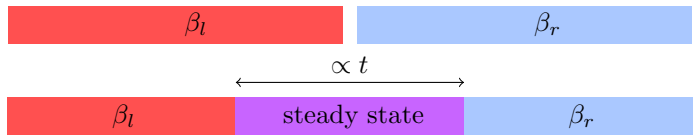

\centering\tikz{
	\fill[light red]
    	(0,0) rectangle(4.4,.5);
    \fill[light blue]
    	(4.6,0) rectangle(9,.5);
    \fill[black]
        (2.2,.25) node[anchor=west]{$\beta_l$}
        (6.8,.25) node[anchor=west]{$\beta_r$};
}
\centering\tikz{
	\fill[light red]
    	(0,0) rectangle(3,.5);
    \fill[lilac]
    	(3,0) rectangle(6,.5);
    \fill[light blue]
    	(6,0) rectangle(9,.5);
    \draw[thin, style= <->]
    	(3,.7)--(6,.7);
    \fill[black]
    	(4.5,.7) node[anchor=south]{$\propto t$}
        (1.25,.25) node[anchor=west]{$\beta_l$}
        (3.5,.25) node[anchor=west]{steady state}
        (7.25,.25) node[anchor=west]{$\beta_r$};
}
\caption{We consider two semi-infinite critical systems, initially thermalized at different temperatures. At time $t=0$, the systems are connected at a point so that energy can flow between them.
After time $t$, there is a sharply defined region of size $R\propto t$ in which there is a steady state description. 
We want to consider the behaviour of the entanglement between the left- and right baths as time evolves. 
}
\label{fig:physicalsituation}
\end{figure}

Let us denote by $H^l$ and $H^r$ the Hamiltonians of the left and right subsystems, respectively. The density matrix describing the initial state, with independently thermalized subsystems, is
\eq{
\rho_0=e^{-\beta_lH^l-\beta_rH^r}.
}
The expectation value taken in the initial state is denoted as
\eq{
\langle\cdots\rangle_0=\frac{\text{Tr}(\rho_0\cdots)}{\text{Tr}(\rho_0)},
}
Note that since $[H^l,H^r]=0$, the expectation values factorise into the left and right systems: if $O_1(x)$ and $O_2(y)$ are local observables at positions $x<0$ and $y>0$, respectively, then
\eq{\label{eq:corr-l-r}
\langle O_1(x)O_2(y)\rangle_0=\langle O_1(x)\rangle_{\lt}\langle O_2(y)\rangle_{\rt}\qquad (x<0,\,y>0),
}
where the expectation values $\langle\cdots\rangle_{\lt/\rt}$ are taken with respect to $e^{-\beta_l H^l}$ and $e^{-\beta_r H^r}$, respectively.

After the quench, a connection is added between the left and right subsystems, and the full Hamiltonian is
\eq{
H=H^l+H^r+\delta H.
}
The term describing the connection $\delta H$ does not commute with either $H^l$ or $H^r$. Although it may have a vanishingly small effect on the value of the total energy, it affects the {\em dynamics} in an important way. The density matrix evolves with time according to the full Hamiltonian, $\rho_0(t) = e^{-iHt}\rho_0 e^{iHt}$, and for any time $t>0$
it does not factorise into left and right subsystems anymore. Expectation values of observables in the state at time $t$ after the quench can naturally be written in terms of expectation values with respect to $\rho_0$ of time-evolved observables,
\eq{\label{rho0tA}
\frac{\text{Tr}(\rho_0(t)O)}{\text{Tr}(\rho_0(t))}=\langle O(t)\rangle_0,
}
with $O(t)=e^{iHt}Oe^{-iHt}$.

At all times $t>0$, and in particular in the non-equilibrium steady state occurring at infinite times, the density matrix corresponds to non-trivial, generically non-thermal mixed states. Hence in order to study the dynamics of the entanglement after the quench and in the steady state in the presence of an energy current, we need to use a measure of entanglement that is appropriate for any mixed states. One such measure is the {\em logarithmic negativity} \cite{VidalWerner2002}. This provides real numbers characterizing the quantity of entanglement between any two subsystems in mixed states. That is, for any decomposition of the Hilbert space as ${\cal H}={\cal H}_{A_1} \otimes {\cal H}_{A_2} \otimes {\cal H}_B$ (where ${\cal H}_B$ may be trivial) and for any density matrix $\rho$ on ${\cal H}$, the logarithmic negativity ${\cal E}_{A_1,A_2}(\rho)$ gives the quantity of entanglement present in $\rho$ between subsystems ${\cal H}_{A_1}$ and ${\cal H}_{A_2}$. The results of \cite{CalabreseCardyTonni2012,CalabreseCardyTonni2013} provide, in field theory, the logarithmic negativity as a certain nontrivial limit on averages with respect to $\rho$ of observables determined by $A_1$ and $A_2$. The observables involve the {\em branch-point twist fields}, which are local observables of the replicated (multi-copy) field theory model \cite{Knizhnik1987,CalabreseCardy2004,CardyCastro-AlvaredoDoyon2007}. Hence we may use \eqref{rho0tA} in order to evaluate the entanglement negativity in field theory.

Below we will study the cases where $A_1 = [u_1,v_1]$ and $A_2=[u_2,v_2]$ represent disjoint, contiguous sets of local degrees of freedom (sites in the quantum chain).

\begin{figure}[h!]
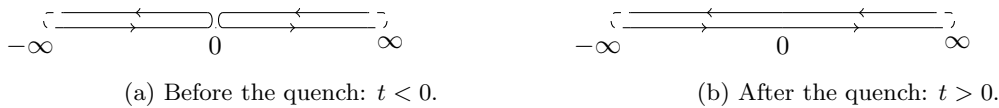

\centering\begin{subfigure}[t]{.5\textwidth}\tikz[anchor=baseline]{
	\begin{scope}[decoration={
    markings,
    mark=at position 0.5 with {\arrow{>}}}]
		\draw[thin,postaction={decorate}]    (2,0.2)
        --(0.1,0.2);
		\draw[thin,postaction={decorate}]    (0.1,0)--(2,0);
		\draw[thin,postaction={decorate}]    (-2,0)--(-0.1,0);
		\draw[thin,postaction={decorate}]    (-0.1,0.2)
        --(-2,0.2);
		\draw[thin]
			(-0.1,0)to[in=0,out=0](-0.1,0.2)
			(0.1,0)to[in=180,out=180](0.1,0.2);
		\draw[thin, dashed]
			(0,0)--(0,0)node[anchor=north]{$0$}
			(-2,0)node[anchor=north east]{$-\infty$}--(-2.2,0)to[in=180,out=180](-2.2,0.2)--(-2,0.2)
			(2,0)node[anchor=north west]{$\infty$}--(2.2,0)to[in=0,out=0](2.2,0.2)--(2,0.2);
	\end{scope}
}
\caption{Before the quench: $t<0$.
}
\end{subfigure}
\begin{subfigure}[t]{.5\textwidth}
\tikz[anchor=baseline]{
	\begin{scope}[decoration={
    markings,
    mark=at position 0.5 with {\arrow{>}}}]
	\draw[thin,postaction={decorate}]    (2,0.2)--(1,0.2)node[anchor=south]{}--(0,0.2);
	\draw[thin,postaction={decorate}]    (0,0)node[anchor=north]{$0$}--(1,0)node[anchor=north]{}--(2,0);
	\draw[thin,postaction={decorate}]    (-2,0)--(-1,0)node[anchor=north]{}--(0,0);
	\draw[thin,postaction={decorate}]    (0,0.2)--(-1,0.2)node[anchor=south]{}--(-2,0.2);
	\draw[thin, dashed]
		(-2,0)node[anchor=north east]{$-\infty$}--(-2.2,0)to[in=180,out=180](-2.2,0.2)--(-2,0.2)
		(2,0)node[anchor=north west]{$\infty$}--(2.2,0)to[in=0,out=0](2.2,0.2)--(2,0.2);
\end{scope}
}
\caption{After the quench: $t>0$.
}
\end{subfigure}
\caption{
Before the time of connection, the systems (at low temperatures) are described by two copies of the same CFT, each on the half-line, with different temperatures. After the connection, the total system is described by a CFT on the line, but one can not associate a temperature to the state.
}
\label{fig:localquench}
\end{figure}

Let us now assume that the quantum system is {\em critical}, and that the dynamical exponent is unity (for instance a quantum chain at a critical point, such as the Heisenberg model). If we assume that the initial temperatures, $\beta_l^{-1}$ and $\beta_r^{-1}$, of the left and right halves are small as compared to microscopic energy scales (for instance, the typical energy of a link in the quantum chain), we may describe the physics by using Conformal Field Theory (CFT). The quench process that we described above has been studied within CFT in \cite{bernard2012energy, BernardDoyon2014}, and as explained there, a current-carrying non-equilibrium steady state develops at large times. From standard CFT arguments, the fields that are in the same conformal family as the energy and momentum density separate into right and left movers. On each semi-infinite initial, separate subsystems, right and left movers are related to each other via conformal boundary conditions at the endpoints, see Figure \ref{fig:localquench}. The effect of the local quench is to modify the dynamics in such a way that at times $t>0$, right and left movers flow continuously through the total system, at the speed of light (the Lieb-Robinson velocity of the quantum chain). This has the effect of producing a light cone, outside which the initial independentaly thermalized states are observed, and inside which a non-equilibrium steady state occurs. The steady state is completely described by independently thermalizing right and left movers at inverse temperatures $\beta_l$ and $\beta_r$, respectively, or equivalently by boosting a thermal state of rest-frame inverse termperature $\sqrt{\beta_l\beta_r}$ and boost velocity $(\beta_r-\beta_l)/(\beta_r+\beta_l)$. See Figure \ref{fig:physicalsituation}.

Below we will combine the CFT description of the quench problem and the emerging steady state with the twist-field expressions for logarithmic negativity in order to study the universal dynamics of entanglement in the presence of energy flows in critical systems.


\begin{figure}[h!]
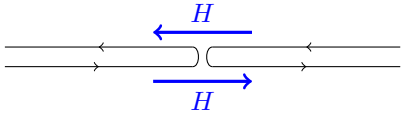

\centering\tikz[scale=1.3,anchor=baseline]{
	\begin{scope}[decoration={
    markings,
    mark=at position 0.5 with {\arrow{>}}}]
		\draw[thin,postaction={decorate}]    (2,0.2)
        --(0.1,0.2);
		\draw[thin,postaction={decorate}]    (0.1,0)--(2,0);
		\draw[thin,postaction={decorate}]    (-2,0)--(-0.1,0);
		\draw[thin,postaction={decorate}]    (-0.1,0.2)
        --(-2,0.2);
		\draw[thin]
			(-0.1,0)to[in=0,out=0](-0.1,0.2)
			(0.1,0)to[in=180,out=180](0.1,0.2);
      	\draw[very thick,color=blue,->]	
        (0.5,0.35)--(0,0.35)node[anchor=south]{$H$}--(-0.5,0.35);
        \draw[very thick,color=blue,->]
        (-0.5,-0.15)--(0,-0.15)node[anchor=north]{$H$}--(0.5,-0.15);
	\end{scope}
}
\caption{
Expectation values at time $t$ after the quench are computed using time evolved observables on the disconnected state. The observables are evolved with the full Hamiltonian $H$, which includes the connection between the left and right systems.  
}
\label{fig:heisenberg-picture}
\end{figure}

{\ }\\
{\bf Remark.} {\em The quantum-chain precursor to branch-point twist fields are cyclic replica permutation operators, studied in \cite{Castro-AlvaredoDoyon2011}. Generalizing the field-theory arguments of \cite{CalabreseCardyTonni2012,CalabreseCardyTonni2013}
and using these quantum-chain operators, one obtains the logarithmic negativity via averages of local observables in quantum chains. Hence, one may also use \eqref{rho0tA} in quantum chains in order to study the dynamics of the negativity dynamics, by replacing the observable $O$ with products of cyclic replica permutation operators instead of branch-point twist fields.}

\subsection{Main results of this paper}
\begin{figure}[h!]
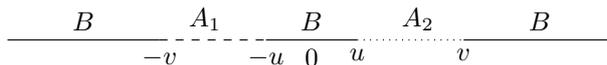

\centering\tikz[scale=2]{
\draw	(-2,0)--(-1.5,0)node[anchor=south]{$B$}--(-1,0)
		(-0.3,0)--(0,0)node[anchor=north]{$0$}node[anchor=south]{$B$}--(0.3,0)
		(1,0)--(1.5,0)node[anchor=south]{$B$}--(2,0);
\draw[dashed]	
	(-1,0)node[anchor=north]{$-v$}--(-0.7,0)node[anchor=south]{$A_1$}--(-0.3,0)node[anchor=north]{$-u$};
\draw[dotted]	(0.3,0)node[anchor=north]{$u$}--(0.7,0)node[anchor=south]{$A_2$}--(1,0)node[anchor=north]{$v$};
}
\caption{The negativity between two parts $A_1=[-v,-u]$ and $A_2=[u,v]$ of finite length, and equal distance from the point of connection. The negativity is a measure for the entanglement between two (not necessarily complementary) regions $A_1$ and $A_2$ of a system $A_1\cup A_2\cup B$.}
\end{figure}

We denote by $\logneg_{A_1,A_2}(t;\beta_l,\beta_r)$ the logarithmic negativity between degrees of freedom lying on subsets $A_1$ and $A_2$, a time $t$ after the connection, with initial left and right inverse temperatures $\beta_l$ and $\beta_r$ respectively (see Figure~\ref{fig:localquench}). In the following, we will denote the logarithmic negativity in equilibrium (i.e. in a system where no quench has taken place) at inverse temperature $\beta$ by $\logneg^{eq}_{A_1,A_2}(\beta)$. We find the following.

For technical reasons, calculations will mainly be restricted to CFT models with trivially factorized pairing between holomorphic and antiholomorphic modules of the CFT (which we will refer to as {\em trivial pairing data}). However, certain results generalize to arbitrary pairing, as we will indicate.

For CFT models with trivial pairing data, the logarithmic negativity between two intervals of equal length, $A_1=[-v,-u]$ and $A_2=[u,v]$ is given by the average of two expressions that each depend only on one of the temperatures,
\eq{\label{eq:logneg_unequal_temp}
\logneg_{[-v,-u],[u,v]}(t;\beta_l,\beta_r)=\frac{1}{2}\left(\logneg_{[-v,-u],[u,v]}(t;\beta_l)+\logneg_{[-v,-u],[u,v]}(t;\beta_r)\right),
}
where $\logneg_{[-v,-u],[u,v]}(t;\beta)$ is the negativity obtained from intially thermalizing both halves at the same temperature. We find different behaviour for the latter function depending on the length $v-u$ of the intervals we are measuring, and the time $t$ after connection.

Entanglement starts building after $t>u$. For intermediate times $u<t<v$, the following relation between the logarithmic negativity after the quench in terms of the equilibrium negativity between different intervals holds for CFT models with trivial pairing data:
\eq{\label{eq:result-neg-u<t<v}
\logneg_{[-v,-u],[u,v]}(t;\beta)
=\logneg^{eq}_{[-v,-u]\cup[t,v],[u,t]}(\beta)+\ln\Ctttsq- \ln c_{1/2} +3\ln g
,\qquad u<t<v.
}
Recall that the equilibrium negativities $\logneg^{eq}_{\tilde{A}_1,\tilde{A}_2}$ are calculated in an infinite system where no quench has taken place; the effect of the quench is encoded in the now changed intervals $\tilde{A}_1=[-v,-u]\cup[t,v]$ and $\tilde{A}_2=[u,t]$. The constant $\Ctttsq$ is the limit $n\rightarrow 1$ from even $n$ of a universal 3-point coupling characteristic of the CFT model, whereas $c_{1/2}$ 
is a non-universal constant that depends on the microscopic details of the quantum chain. Finally, we have a term which is a multiple of $\ln g$, the boundary entropy \cite{AffleckLudwig1991}. 
For late times, $t>v$, the observables measuring the negativity are in the NESS. The logarithmic negativity in the NESS does not depend on time, and the relation between the negativity after the quench and equilibrium expressions simplifies:
\eq{\label{eq:result-neg-t>v}
\logneg_{[-v,-u],[u,v]}(t;\beta)=\logneg^{eq}_{[-v,-u],[u,v]}(\beta),\qquad t>v.
}
Note that in this regime the above relation~\eqref{eq:result-neg-t>v} and the relation~\eqref{eq:logneg_unequal_temp} are consistent with the state being a boosted thermal state with boost velocity $(\beta_r-\beta_l)/(\beta_r+\beta_l)$ and rest-frame inverse temperature $\sqrt{\beta_l \beta_r}$. This latter descrition is expected to hold for CFTs with nontrivial pairing data as well.
\begin{figure}[h!]
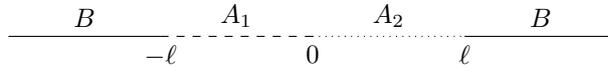

\centering\tikz[scale=2]{
\draw	(-2,0)--(-1.5,0)node[anchor=south]{$B$}--(-1,0)
		(1,0)--(1.5,0)node[anchor=south]{$B$}--(2,0);
\draw[dashed]	
	(-1,0)node[anchor=north]{$-\ell$}--(-0.5,0)node[anchor=south]{$A_1$}--(0,0);
\draw[dotted]	(0,0)node[anchor=north]{$0$}--(0.5,0)node[anchor=south]{$A_2$}--(1,0)node[anchor=north]{$\ell$};
}
\caption{Negativity between two parts $A_1$ and $A_2$ of finite length $\ell$. }
\end{figure}

The equilibrium expressions for the negativity in the above relations generally depend on the CFT model in question, and therefore an explicit solution cannot be found from CFT methods. However, specializing to the case $A_1=[-\ell,0]$ and $A_2=[0,\ell]$, we may find approximate solutions for certain limits of $\ell$ and $t$.

For instance, just after the quench, the logarithmic negativity calculated using \eqref{eq:result-neg-u<t<v}, with $u=0$ and $v=\ell$, becomes
\eq{\label{eq:logneg_0<t<ell}
\logneg_{[-\ell,0],[0,\ell]}(t;\beta)=\frac{c}{2}\ln\frac{t}{\delta}+\ln\Ctttsq+ \ln c_{1/2} +3\ln g,
\qquad t\ll\text{ any other scale},
}
where $\delta$ is a non-universal factor related to the lattice spacing of the underlying quantum chain and $c$ is the central charge of the CFT model. Note that the behaviour just after the quench does not depend on the temperatures of the systems before the quench.

Another limit we can take is $\ell\gg t\rightarrow\infty$. In this limit, the equilibrium terms in \eqref{eq:result-neg-u<t<v}, again specializing to $u=0$ and $v=\ell$, can be found to be of the form
\eq{\label{eq:logneg_0<t<inf}
\lim_{s\rightarrow\infty}\logneg_{[-\infty,0],[0,\infty]}(s;\beta)=\frac{c}{2}\ln\frac{\beta}{2\pi\delta}
+3\ln\Ctttsq+ \ln c_{1/2} 
+3\ln g.
}
This indicates a plateau, which is different from the plateau reached in the NESS. One can interpret the second term as arising from the fact that the entanglement builds up around the two boundary points, which in this limit are far away from each other.

We expect the asymptotic result for small $t$ \eqref{eq:logneg_0<t<ell} with \eqref{eq:logneg_unequal_temp} to hold for more general module pairing in the CFT, but in the ``prethermal" regime, the result \eqref{eq:logneg_0<t<inf} may have corrections for CFTs with nontrivial pairing data.
The difference between the logarithmic negativity just after the quench and in the limit $\ell>t\rightarrow\infty$ is a universal function of $t/\beta$:
\eq{
\logneg_{[-\infty,0],[0,\infty]}(t;\beta)-\lim_{s\rightarrow\infty}\logneg_{[-\infty,0],[0,\infty]}(s;\beta)=\frac{c}{2}\ln\frac{2\pi t}{\beta}
-2\ln\Ctttsq\qquad t\ll\text{ any other scale}.
}

The logarithmic negativity in the NESS (i.e. for $t>\ell$) does not depend on time, and is given by
\eq{
	\label{negNESS}
\logneg^{NESS}_{[-\ell,0],[0,\ell]}(\beta)=\logneg_{[-\ell,0],[0,\ell]}(t>\ell;\beta)=\frac{c}{4}\ln\left(\frac{\beta}{2\pi\delta}\tanh\frac{\pi\ell}{\beta}\right)+\ln\Ctttsq +\ln c_{1/2}.
}
In this regime it is expected that this result does not depend on pairing data.


%
\subsection{Comparison to results in the literature}

The CFT corresponding to the harmonic chain numerically studied in \cite{EislerZimboras2014b} has trivially factorized module pairing. Therefore, all above results should apply to this case. Using general CFT arguments and their numerical results, the authors of \cite{EislerZimboras2014b} conjectured an expression for the logarithmic negativity in the nonequilibrium steady state (NESS). Our result \eqref{negNESS} at $c=1$ confirms that this conjecture is correct. Further, other numerical results found in \cite{EislerZimboras2014b} for the regime before the system reaches the steady state suggest that the logarithmic negativity builds up quickly (logarithmically), and then saturates to a finite value, before hitting the NESS regime in which it saturates at a lower value. These general features are in agreement with the above results.

%
%


\section{Branch-point twist fields and a real-time CFT approach to entanglement dynamics}
\label{sec:twistfields}

Some measures of entanglement, such as the von Neumann and R\'{e}nyi entropies, and the (logarithmic) negativity, can be expressed using the replica trick in terms of correlation functions of so-called \textit{branch-point twist fields} \cite{Knizhnik1987,CalabreseCardy2004,CardyCastro-AlvaredoDoyon2007}, associated to the permutation symmetry of the copies. Branch-point twist fields exist in any replica, $n$-copy QFT model, and are associated with the symmetry under permutation of the copies. The main property of the twist field of interest, associated with a cyclic permutation, is the exchange property
\begin{subequations}\label{eq:twistfields}
\eq{
\varphi_i(y,t)\,\twist(x,t)=\left\{
\begin{array}{ll}
	\twist(x,t)\,\varphi_{i}(y,t)&(y<x) \\ \twist(x,t)\,\varphi_{i+1}(y,t)\qquad&(x<y).
\end{array}
\right.
}
Similarly, the ``anti-twist'' field, associated with the inverse cyclic permutation, satisfies
\eq{
\varphi_i(y,t)\,\atwist(x,t)=\left\{
\begin{array}{ll}
	\atwist(x,t)\,\varphi_{i}(y,t)&(y<x) \\ \atwist(x,t)\,\varphi_{i-1}(y,t)\qquad&(x<y).
\end{array}
\right.
}
\end{subequations}
The twist fields are local, primary fields, and their scaling dimension was found in \cite{Knizhnik1987,CalabreseCardy2004} to be 
\eq{\label{eq:twist-dimension}
d_n=\frac{c}{12}(n-n^{-1}).
}
We will use the CFT normalization
\eq{\label{eq:normalization}
\twist(x)\atwist(y)\sim (x-y)^{-2d_n}.
}

In general, the trace of the product of reduced density matrices appearing in the expression for the R\'{e}nyi entropy $S_A^{(n)}$ for a region $A$ consisting of $N$ cuts $A=[u_1,v_1]\cup\cdots\cup[u_N,v_N]$, is given, for an initial state represented by $\langle\cdots\rangle$, by the following $2N$-point function
\eq{\label{eq:EE_disjoint}
\text{Tr}\rho_A^n=c_n^N\delta^{2Nd_n}\langle\twist(u_1)\atwist(v_1)\cdots\twist(u_N)\atwist(v_N)\rangle,
}
where $\sqrt{c_n}$ is a nonuniversal constant encoding the conical singularity at the positions of the twist fields (this depends on the number of sheets) \footnote{Note the different way in which our constant $c_n$ appears in the formulae compared to \cite{CalabreseCardy2004} and papers after that: in our case, the constant appears as a pair of twist fields is inserted, whereas in other works it appears in the computation of the correlation function of twist fields.}, and $\delta$ is a short-distance regulator such as a lattice spacing. Note that the expression on the right-hand side of \eqref{eq:EE_disjoint} is dimensionless, as the twist fields have dimension $d_n$.

For the negativity, different configurations of twist fields are required. For instance, dividing a system into three subsystems $A_1$, $A_2$ and $B$, consider the negativity measuring of the entanglement between  $A_1$ and $A_2$, disregarding the entanglement with the third subsystem $B$.

\begin{figure}[h!]
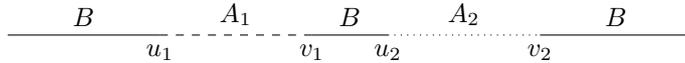

\centering\tikz{
\draw[thin]	(0,0)--(1,0)node[anchor=south]{$B$}--(2,0);
\draw[thin]	(4,0)--(4.5,0)node[anchor=south]{$B$}--(5,0);
\draw[thin]	(7,0)--(8,0)node[anchor=south]{$B$}--(9,0);
\draw[thin,dashed]	(2,0)node[anchor=north]{$u_1$}--(3,0)node[anchor=south]{$A_1$}--(4,0)node[anchor=north]{$v_1$};
\draw[thin,dotted]	(5,0)node[anchor=north]{$u_2$}--(6,0)node[anchor=south]{$A_2$}--(7,0)node[anchor=north]{$v_2$};
}
\caption{A picture representing the negativity between part $A_1=[u_1,v_1]$ (dashed line) and $A_2=[u_2,v_2]$ (dotted line).}
\end{figure}

In a state represented by $\langle\cdots\rangle$, the logarithmic negativity between the systems $A_1=[u_1,v_1]$ and $A_2=[u_2,v_2]$ (with $A=A_1\cup A_2$ and $T_2$ denotes partial transposition with respect to $A_2$) is given by \cite{CalabreseCardyTonni2012,CalabreseCardyTonni2013}
\eq{\label{eq:logneg}
\logneg_{A_1,A_2}=\lim_{n\rightarrow 1}\log\text{Tr}|\rho_A^{T_2}|^n=\lim_{\substack{n\rightarrow 1 \\ n\text{ even}}}\log\text{Tr}(\rho_A^{T_2})^{n}=\lim_{\substack{n\rightarrow 1 \\ n\text{ even}}}\log c_n^2\delta^{4d_n}\langle\twist(u_1)\atwist(v_1)\atwist(u_2)\twist(v_2) \rangle,
}
where we note that the last two equalities hold only when taking the limit $n\rightarrow 1$ analytically continuing from an even number of copies.

Note that in the expressions for the R\'{e}nyi entropies \eqref{eq:EE_disjoint} and the logarithmic negativity \eqref{eq:logneg} we have not specified the state the system is in. In the following, we will consider the local quench as described in section~\ref{sec:localquench}. Our density matrices will therefore have the following dependencies
\eq{
\rho_A=\rho_A(t;\beta_l,\beta_r).
}
In some cases we will find relations between the negativity after the quench and negativities in equilibrium, where a system is thermalized at a certain inverse temperature $\beta$. We will denote the equilibrium expressions with
\eq{
\rho^{eq}_A=\rho^{eq}_A(\beta).
}

\subsection{Chiral twist fields}
\label{subsec:chiral-twistfields}
%
In CFT, local fields decompose into local holomorphic (or chiral) and anti-holomorphic (or anti-chiral) components. Seen as generating Virasoro modules, local fields may be written as $\varphi(x)=\varphi^+(x)\varphi^-(x)$ (more precisely, there is a basis such that this holds). Under homogeneous time evolution, a local field $\varphi(x,t)$ evolves as $\varphi(x,t)=\varphi^+(x-t)\varphi^-(x+t)$, again in the sense of generators for Virasoro modules, and thus time evolution separates the components. The field-theory meaning of this decomposition as modules is that the stress-tensor has independent integer-power ``holomorphic'' (i.e. as function of $x-t$) and ``anti-holomorphic'' (i.e. as function of $x+t$) expansions with finite-order singularities. This implies that each component $\varphi^+(x)$ and $\varphi^-(x)$ commutes with energy and momentum densities at space-like distances -- thus fulfilling the requirement of locality. Using this decoupling, we can formally define chiral branch point twist fields,
which are defined in such a way that they cyclically permute only chiral or anti-chiral components.
\begin{subequations}
For example, the equal-time exchange relations for the right-moving branch-point twist field with a right-moving field $\varphi^{+}$ is
\eq{
\varphi^{+}_i(y)\twist^{+}(x)=\left\{
\begin{array}{ll}
	\twist^{+}(x)\,\varphi^{+}_{i}(y) & (y<x) \\ 
    \twist^{+}(x)\,\varphi^{+}_{i+1}(y)\qquad & (x<y)
\end{array}
\right.
}
while the equal-time exchange relation with a left-moving field $\varphi^{-}$ is simply
\eq{
\varphi^{-}_i(y)\twist^{+}(x)=\twist^{+}(x)\varphi^{-}_i(y).
}
Similar relations hold for left-moving twist fields and for the anti-twist fields.
\end{subequations}
By considering exchange relations with the stress-energy tensor, it can immediately be seen that these chiral twist fields commute with the full energy and momentum densities of the replica ($n$-copy) theory. Hence, they are local fields.

In fact, one can infer from the above that these chiral twist fields must be related to the usual twist fields via its own holomorphic factorization, $\twist(x)=\twist^{+}(x)\twist^{-}(x)$. In particular, their conformal dimension is given by
\eq{\label{eq:confdim}
\cdim=\frac{c}{24}\left(n-\frac{1}{n}\right).
}
Note that the chiral twist fields carry spin, since for a chiral twist field the difference between its holomorphic dimension and its anti-holomorphic dimension is $\Delta\neq 0$ in general.

When chiral twist fields are brought close to each other, they may or may not have a divergence, depending on the chirality of the fields. Two fields of different chirality do not produce a divergence. For instance,
\eq{
\twist^{+}(x)\twist^{-}(y)\stackrel{x\rightarrow y}{\sim}\twist^{+}(y)\twist^{-}(y)=\twist(y).
}
However, fields of the same chirality do produce a divergence, for instance in the following OPE,
\eq{
\twist^{+}(x)\twist^{+}(y)\stackrel{x\rightarrow y}{\sim}(x-y)^{\cdimsq-2\cdim}(\twist^{+})^2(y)C_{\twist^{+}\twist^{+}}^{(\twist^{+})^2},
}
where the structure constant $C_{\twist^{+}\twist^{+}}^{(\twist^{+})^2}$ is a property of the CFT model under consideration, and the conformal dimension of the field $(\twist^{+})^2$, which is equal to the conformal dimension of $(\twist^{-})^2$, $(\atwist^{+})^2$ and $(\atwist^{-})^2$, is given by
\eq{\label{eq:cdimsq}
\cdimsq:=\left\{
\begin{array}{ll}
\cdim & n \text{ odd}\\
2\cdim[n/2] & n \text{ even}
\end{array}
\right.
}
Similarly, we have the OPE
\eq{\label{eq:ope-t-at}
\twist^{+}(x)\atwist^{+}(y)\stackrel{x\rightarrow y}{\sim}(x-y)^{-2\cdim},
}
where the normalization in \eqref{eq:normalization} was used.

\subsection{Pairings of holomorphic and anti-holomorphic fields}

It is well known that, although holomorphic factorization in CFT is true at the level of Virasoro representations -- that is, the factors are local fields -- it does not hold, generically, at the level of the operator algebra. That is, the OPE between fields $\varphi_i(x)$ and $\varphi_j(y)$ is generically not the product of the OPEs between their individual holomorphic, $\varphi_i^+(x)\varphi_j^+(y)$, and anti-holomorphic, $\varphi_i^-(x)\varphi_j^-(y)$, components. For instance, one may have the diagonal structure (here for spinless fields)
\eq{
	\varphi_i(x)\varphi_j(y) = \sum_k
	C_{ij}^k \bar C_{ij}^k (x-y)^{d_k-d_i-d_j} \varphi_k(y)
}
instead of the factorized structure
\eq{
	\varphi_i(x)\varphi_j(y) = \sum_{k,k'}
	C_{ij}^k \bar C_{ij}^{k'} (x-y)^{\Delta_k^+ + \Delta_{k'}^--d_i-d_j} \varphi_k^+(y)\varphi^-_{k'}(y).
}
One may say that although holomorphic and anti-holomorphic components are local, they generically have semi-local properties with respect to the operator algebra. This generically affects, in particular, the chiral twist fields that we introduced above.

It is known that the particular structure of the OPEs constitute additional data of the CFT model under consideration, which, along with the central charge, the set of modules involved and the chiral OPE coefficients, fully characterize the CFT model. This additional data may be referred to as the pairing data of the model. Because of the separation between the chiral and anti-chiral components of the stress-tensor (and of other symmetry currents), highest-weight modules always appear, in any OPE, in a factorized fashion, hence the only pairing data necessary is that identifying the pairing between modules.

The pairing of holomorphic and anti-holomorphic components is a manifestation, at the CFT level, of the fact that, at the quantum-chain level, time-evolved fields are not in general locally supported on end-points of the light-cone, but rather are supported on the full interval lying inside the light-cone. Formally, we may represent this pairing as a connection between holomorphic and anti-holomorphic components, and this connection constrains the OPEs involving the separate components.

Models in which the pairing is trivially factorized are those that are completely built out of symmetry currents: those where all representations involved are the representations associated to the symmetry algebra itself. Free-boson (harmonic chains) and free-fermion models display this property. By construction, in a model with trivially factorized pairing, the OPEs of twist fields in the replica model also trivially factorizes.

It is beyond the scope of this paper to go into any detail of the effect of the pairing data on the dynamics of entanglement. Below we make general comments on this,  
but mostly consider the spacial case of factorized pairing.
However, many of our results, we expect, do not depend on the pairing data.

{\ }\\
{\bf Remark.} {\em The pairing data of a CFT model affects $n$-point function calculations for $n\geq 4$; these can be calculated generically by inserting OPEs, and pairing data is necessary within this procedure. Pairing data leads to expressions of 4-point functions as particular linear combinations of products of conformal blocks. One can see, on the other hand, that 2- and 3-point functions do not depend on the pairing data.}

\subsection{Dynamics after a local quench}

The definition of the chiral twist fields is invariant under conjugation by any unitary operator. Considering the time-evolution operator, this allows us to deduce the behaviour of branch-point twist fields under time evolution. For example,  evolving with the Hamiltonian $H$ after connection, we get the following relation for right-moving fields:
\eq{
\varphi^{+}_i(\tilde{y})\,e^{-iHt}\twist^{+}(x)e^{iHt}=\left\{
\begin{array}{ll}
	e^{-iHt}\twist^{+}(x)e^{iHt}\,\varphi^{+}_{i}(\tilde{y})&(\tilde{y}<x-t) \\ e^{-iHt}\twist^{+}(x)e^{iHt}\,\varphi^{+}_{i+1}(\tilde{y})\qquad&(x-t<\tilde{y})
\end{array}
\right.
}%
where $\tilde{y}:=y-t$. From this relation one can see that under time evolution with the Hamiltonian after connection, the right-moving twist field indeed evolves as $e^{-iHt}\twist[]^{+}(x)e^{iHt}=\twist[]^{+}(x-t)$, and the left-moving fields as $e^{-iHt}\twist[]^{-}(x)e^{iHt}=\twist[]^{-}(x+t)$. The same relations hold for the other twist fields defined in \eqref{eq:twistfields}.

After time evolution, chiral twist fields are evaluated in the disconnected state $\langle\cdots\rangle_0$ defined by \eqref{eq:corr-l-r}. It may be convenient to define a so-called unfolding map, in which left- and right moving fields of the half line are mapped to holomorphic fields on the full line, as follows:
\begin{subequations}
\eq{\label{eq:map-twist}
\twist^+(x)\mapsto\chtwist(x),\qquad \twist^{-}(x)\mapsto\chatwist(-x),\qquad \atwist^+(x)\mapsto\chatwist(x),\qquad \atwist^{-}(x)\mapsto\chtwist(-x),
}
The holomorphic twist fields $\chtwist$ and $\chatwist$ are defined on the unfolded line, and will be evaluated in thermal states as per
\eq{\label{eq:map-corr}
\langle\cdots\rangle^{\lt/\rt}\mapsto\langle\cdots\rangle^{ch}_{\beta_{\lt\rt}},
}\label{eq:unfoldingmap}
\end{subequations}
where $\langle\cdots\rangle^{ch}_\beta$ denotes the thermal expectation value taken in the holomorphic sector. The conformal dimension of these holomorphic twist fields $\tau$ and $\tilde{\tau}$ are again given by \eqref{eq:confdim}. Similar relations hold for the anti-twist fields.

Because of the generically nontrivial pairing data of the CFT, after time evolution, where some chiral components have evolved through the origin $x=0$ and changed side, the expectation value in the state $\langle\cdots\rangle_0$ does not generically factorize into a product of expectation values in $\langle\cdots\rangle^{\lt/\rt}$. Indeed, pairing may imply connections between components that are positioned in different halves of the system. Hence, the holomorphic expectation values $\langle\dots\rangle^{ch}_{\beta_{\lt\rt}}$ occurring after unfolding should be understood as thermal conformal blocks, and the time-evolve correlation function is a sum of products of such blocks.

However, with factorized pairing, even after time evolution, the expectation value of chiral components in $\langle\cdots\rangle_0$ does factorize into a product expectation values in $\langle\cdots\rangle^{\lt/\rt}$. Further, in this case, any thermal expectation value of full (holomorphic times anti-holormophic) fields also factorizes into its chiral component, so that we have a convenient relation between the holomorphic twist fields in the unfolded system and the twist fields on the line:
\eq{\label{eq:holomorphic-full-twist}
\langle\chtwist(x)\dots\chatwist(y)\rangle^{ch}=\left(\langle\twist(x)\dots\atwist(y)\rangle\right)^{1/2}.
}

Using \eqref{rho0tA}, \eqref{eq:EE_disjoint} and \eqref{eq:logneg}, one can use these chiral twist fields to describe the dynamics of entanglement after the quench by evolving the fields under the full Hamiltonian $H$ and then considering the time evolved fields in the disconnected system. This means  all twist fields that after time evolution are to the left of the defect should be evaluated on the half-line at temperature $\beta_l$, and all twist fields that are to the right of the defect, should be evaluated on the half-line at temperature $\beta_r$. These expressions can be simplified by unfolding the left- and right systems, and one is left with products of two holomorphic expressions (conformal blocks) on the line, evaluated at different temperatures. However, additional subtleties arise when the branch cuts emanating from the twist fields cross the point $x=0$ separating the left and right subsystems.

\subsection{Evolution of the entanglement entropy}

As a simple example, take the entanglement entropy between a finite region $A=[u,v]$ and the rest. For reasons of simplicity, we will assume that both $u$ and $v$ are positive. Naively, we have
\eq{
\text{Tr}\rho_A^n(t)=c_n\delta^{2d_n}\langle\twist(u,t)\atwist(v,t)\rangle_0,
}
where $\langle\dots\rangle_0$ is the expectation value taken in the disconnected system and the time-evolved fields are as above. However, this formula might involve non-universal singularities, as becomes clear when the correlation function is expressed in terms of the chiral twist fields:
$\langle\twist^{+}(u-t,0)\twist^{-}(u+t,0)\atwist^{+}(v-t,0)\atwist^{-}(v+t)\rangle_0$. Here we have two different cuts for the left-moving fields and the right-moving fields: $A^{-}=[u+t,v+t]$ and $A^{+}=[u-t,v-t]$, and a subtlety arises when one of the points of these regions crosses the defect. The fact that at the time $t=0$ the boundary conditions at the point of the defect are changed, means that each cut extending across the defect is divided into two shorter cuts, one on each side of the defect. In our calculation of the EE, this is expressed by the insertion of an extra pair of twist fields, giving rise to divergencies that must be regularized.

To make this precise, we consider the EE at the time of the quench as the following limit,
\eq{
\text{Tr}\rho_A^n(0)=c_n\delta^{2d_n}\langle\twist(u,0)\atwist(v,0)\rangle_0=c_n\delta^{2d_n}\lim_{\varepsilon\rightarrow 0}(2\varepsilon)^{2d_n}\langle\twist(u,0)\atwist(t-\varepsilon,0)\twist(t+\varepsilon,0)\atwist(v,0)\rangle_0,
}
where $u<t<v$.
Evolving this over a time $s$ with the connected Hamiltonian $H$, this becomes
\eq{
\begin{split}
\text{Tr}\rho_A^n(s)
&=c_n\delta^{2d_n}\lim_{\varepsilon\rightarrow 0}(2\varepsilon)^{2d_n}\langle\twist(u,s)\atwist(t-\varepsilon,s)\twist(t+\varepsilon,s)\atwist(v,s)\rangle_0\\
&=c_n\delta^{4\cdim}\lim_{\varepsilon\rightarrow 0}(2\varepsilon)^{4\cdim}\langle\twist^{+}(u-s,0)\twist^{-}(u+s,0)\atwist^{+}(t-s-\varepsilon,0)\atwist^{-}(t+s-\varepsilon,0)\\
&\qquad\qquad\twist^{+}(t-s+\varepsilon,0)\twist^{-}(t+s+\varepsilon,0)\atwist^{+}(v-s,0)\atwist^{-}(v+s,0)\rangle_0.
\end{split}
}
The expression after time evolution over a time $s=t$ with $u<t<v$ can be written as a product of expectation values for the left and the right system, with the division at $0$, as in \eqref{eq:corr-l-r}. We can use \eqref{eq:corr-l-r} and the OPE $\twist^{-}(2t-\varepsilon)\atwist^{-}(2t+\varepsilon)\stackrel{\varepsilon\rightarrow 0}{\sim}(2\varepsilon)^{-2\cdim}$ to obtain the expression
\begin{multline}\label{eqqq}
\text{Tr}\rho_A^n(u<t<v)\\
=c_n\delta^{4\cdim}\lim_{\varepsilon\rightarrow 0}(2\varepsilon)^{2\cdim}\langle\twist^{+}(u-t)\atwist^{+}(-\varepsilon)\rangle_{\lt}\langle\twist^{-}(u+t)\twist^{+}(\varepsilon)\atwist^{+}(v-t)\atwist^{-}(v+t)\rangle_{\rt}.
\end{multline}

\begin{figure}[h!]
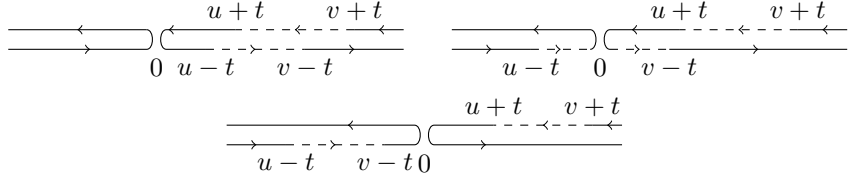

\centering
\tikz[scale=1.3,anchor=base,baseline]{
\begin{scope}[decoration={
    markings,
    mark=at position 0.5 with {\arrow{>}}}]
\draw[thin,postaction={decorate}]	(2.5,0.2)--(2,0.2);
\draw[thin,dashed,postaction={decorate}]    (2,0.2)node[anchor=south]{$v+t$}--(0.8,0.2)node[anchor=south]{$u+t$};
\draw[thin,postaction={decorate}]	(0.8,0.2)--(0.1,0.2)to[in=180,out=180](0.1,0)--(0.5,0);
\draw[thin,dashed,postaction={decorate}]    (0.5,0)node[anchor=north]{$u-t$}--(1.5,0)node[anchor=north]{$v-t$};
\draw[thin,postaction={decorate}]	(1.5,0)--(2.5,0);
\draw[thin,postaction={decorate}]    (-1.5,0)--(-0.1,0)to[in=0,out=0](-0.1,0.2);
\draw[thin,postaction={decorate}]	(-0.1,0.2)--(-1.5,0.2);
\draw[thick]
(0,0)node[anchor=north]{$0$};
\end{scope}
}\hspace{0.5cm}
\tikz[scale=1.3,anchor=base,baseline]{
\begin{scope}[decoration={
    markings,
    mark=at position 0.5 with {\arrow{>}}}]
\draw[thin,postaction={decorate}]	(2.5,0.2)--(2,0.2);
\draw[thin,dashed,postaction={decorate}]    (2,0.2)node[anchor=south]{$v+t$}--(0.8,0.2)node[anchor=south]{$u+t$};
\draw[thin,postaction={decorate}]	(0.8,0.2)--(0.1,0.2)to[in=180,out=180](0.1,0);
\draw[thin,dashed,postaction={decorate}]    (0.1,0)--(0.7,0)node[anchor=north]{$v-t$};
\draw[thin,postaction={decorate}]	(0.7,0)--(2.5,0);
\draw[thin,postaction={decorate}]    (-1.5,0)--(-0.7,0);
\draw[thin,dashed,postaction={decorate}]    (-0.7,0)node[anchor=north]{$u-t$}--(-0.1,0);
\draw[thin,postaction={decorate}]	(-0.1,0)to[in=0,out=0](-0.1,0.2)--(-1.5,0.2);
\draw[thick]
(0,0)node[anchor=north]{$0$};
\end{scope}
}\hspace{0.5cm}
\tikz[scale=1.3,anchor=base,baseline]{
\begin{scope}[decoration={
    markings,
    mark=at position 0.5 with {\arrow{>}}}]
    \draw[thin,postaction={decorate}]    (-2,0)--(-1.4,0);
\draw[thin,dashed,postaction={decorate}]    (-1.4,0)node[anchor=north]{$u-t$}--(-0.4,0)node[anchor=north]{$v-t$};
\draw[thin,postaction={decorate}]	(-0.4,0)--(-0.1,0)to[in=0,out=0](-0.1,0.2)--(-2,0.2);
\draw[thin,postaction={decorate}]	(2,0.2)--(1.7,0.2);
\draw[thin,dashed,postaction={decorate}]    (1.7,0.2)node[anchor=south]{$v+t$}--(0.7,0.2)node[anchor=south]{$u+t$};
\draw[thin,postaction={decorate}]	(0.7,0.2)--(0.1,0.2)to[in=180,out=180](0.1,0)--(2,0);
\draw[thick]
(0,0)node[anchor=north]{$0$};
\end{scope}
}
\caption{EE between a part $A=[u,v]$ (with $u>0$ and $v>0$) and the rest, a time $t$ after connection, with $t<u<v$ (top left), $u<t<v$ (top right) and $u<v<t$ (bottom).}
\label{fig:EE_t}
\end{figure}

As will become clear below, the expectation values in \eqref{eqqq} are regular as $\varepsilon\to0$. Hence we must set the remaining factor $2\varepsilon$ proportional to the short-distance cutoff $\delta$. This is equivalent to reversing the limits of $\varepsilon\rightarrow 0$ and the scaling limit $\delta\rightarrow 0$. The constant of proportionality will generally depend on the number of sheets $n$,
\eq{\label{eq:b_n}
b_n:=\frac{2\varepsilon}{\delta}.
}

The first expectation value in \eqref{eqqq}, corresponding to the left subsystem, can be evaluated by mapping to a chiral theory on the line via the unfolding map \eqref{eq:unfoldingmap}. We obtain
\eq{\label{ttch}
\langle\twist^{+}(u-t)\atwist^{+}(0)\rangle_{\lt}=\langle\chtwist(u)\chatwist(t)\rangle^{ch}_{\beta_l},
}
where the expectation value is taken on the line at inverse temperature $\beta_l$, and we have used translation invariance to shift the rhs expression over $t$. Using the relation \eqref{eq:holomorphic-full-twist} between the holomorphic twist fields and the full twist fields, we can relate this holomorphic expressions to a R\'{e}nyi entropy of a different interval, in a system in equilibrium at a different temperature.

On the other hand, the second expectation value in \eqref{eqqq}, corresponding to the right subsystem, can be re-written as
\eq{\label{rewr}
	\langle\twist^{-}(u+t)\twist^{+}(0)\atwist^{+}(v-t)\atwist^{-}(v+t)\rangle_{\rt} =
	\langle\twist\left(\frac{u+t}2,\frac{u+t}2\right)\atwist(v,t)\rangle_{\rt}.
}
A physical interpretation as a R\'enyi entropy may be obtained by going to a Lorentz boosted frame such that both twist fields, in this frame, are evaluated on the same time slice. In this frame, the state represents a steady state with a thermal flow and with a moving boundary (intersecting the origin of space time), and the R\'enyi entropy is evaluated instantaneously. The boost velocity is $(t-u)/(u+t-2v)$, the resulting interval length is $D=\sqrt{ut-(u+t-2v)v/2}$, and the left-hand side of the interval is at space-time position given by $x_0=t_0=(u+t)(v-t)/(2D)$. Note that the boost velocity is zero at $t=u$ and is the speed of light at $t=v$.

Hence, we find that the R\'{e}nyi entropy of an interval $A=[u,v]$ an intermediate time $u<t<v$ after a quench can be written in terms of equilibrium and boosted-equilibrium quantities as
\eq{\label{eq:renyi_u<t<v}
S^{(n)}_{[u,v]}(u<t<v;\beta_l,\beta_r)
=\frac{1}{2}S^{(n),eq}_{[u,t]}(\beta_l)+
\frac{1}{2}S^{(n),boost}(\beta_r)-\frac{c'_n}{2}+\frac{d_n}{1-n}\ln b_n,
}
where we defined $c'_n:=\ln c_n/(1-n)$, where $S^{(n),eq}_{C}(\beta)$ denotes the R\'{e}nyi entropy between a subsystem $C$ and the rest in the system consisting of the full line, in equilibrium at inverse temperature $\beta$, and where $S^{(n),boost}(\beta_r)$ denotes the R\'enyi entropy in the boosted state described above (which depends on $u$, $v$ and $t$).

A simplification occurs in models with factorized pairing. The unfolding map gives
\eq{
\langle\twist^{-}(u+t)\twist^{+}(\varepsilon)\atwist^{+}(v-t)\atwist^{-}(v+t)\rangle_{\rt}
=\langle\chtwist(-v)\chatwist(-u)\chtwist(t+\varepsilon)\chatwist(v)\rangle^{ch}_{\beta_r},
}
where again we used translation invariance to shift with $t$.
In factorized pairing models, the 4-point function of holomorphic twist fields has a direct interpretation as a R\'enyi entropy, using the relation \eqref{eq:holomorphic-full-twist}. From this we find $S^{(n),boost}(\beta_r) = S^{(n),eq}_{[-v,-u]\cup[t,v]}(\beta_r)$, giving
\eq{\label{eq:renyi_u<t<v2}
S^{(n)}_{[u,v]}(u<t<v;\beta_l,\beta_r)
=\frac{1}{2}S^{(n),eq}_{[u,t]}(\beta_l)+
\frac{1}{2}S^{(n),eq}_{[-v,-u]\cup[t,v]}(\beta_r)-\frac{c'_n}{2}+\frac{d_n}{1-n}\ln b_n.
}
The time dependence is now fully encoded in the intervals $A_j$ in $S^{(n),eq}_{A_j}(\beta)$, with $A_1=[u,t]$ and $A_2=[-v,-u]\cup[t,v]$ for $\beta=\beta_{l,r}$ respectively.

The EE is obtained by taking the limit $n\rightarrow 1$, resulting in the expression
\eq{\label{eq:EE_u<t<v}
S_{[u,v]}(u<t<v;\beta_l,\beta_r)
=\frac{1}{2}S^{eq}_{[u,t]}(\beta_l)+
\frac{1}{2}S^{boost}(\beta_r) -\frac{c'_1}{2}-\frac{c}{12}\ln b_1,
}
and, with factorized pairing,
\eq{\label{eq:EE_u<t<v2}
S_{[u,v]}(u<t<v;\beta_l,\beta_r)
=\frac{1}{2}S^{eq}_{[u,t]}(\beta_l)+
\frac{1}{2}S^{eq}_{[-v,-u]\cup[t,v]}(\beta_r) -\frac{c'_1}{2}-\frac{c}{12}\ln b_1,
}
where we denote with $b_1$ the limit $\lim_{n\rightarrow 1}b_n$\footnote{Note that we need to specify the value of $b_1$ as a limit, as for $n=1$ the twist operators are just the identity operator, and do not depend on position, wherefore the exchange of limits $\delta\rightarrow 0$ and $\varepsilon\rightarrow 0$ works for any $b_1$.}. The last term is equal to the boundary entropy \cite{AffleckLudwig1991}; see Appendix~\ref{sec:boundary_entropy}.

Finally, from Figure~\ref{fig:EE_t}, it is clear that for late times $t>v$, the cuts do not extend across the defect, and we may simply write
\eq{\label{split}
\begin{split}
\text{Tr}\rho_A^n(t>v)
&=c_n\delta^{4\cdim}\langle\twist^{+}(u-t,0)\atwist^{+}(v-t,0)\rangle_{\lt}\langle\twist^{-}(u+t,0)\atwist^{-}(v+t,0)\rangle_{\rt}\\
&=c_n\delta^{4\cdim}\langle\chtwist(u-t)\chatwist(v-t)\rangle^{ch}_{\beta_l}\langle\chtwist(-v-t)\chatwist(u-t)\rangle^{ch}_{\beta_r},
\end{split}
}
which results in the following time-independent expression,
 \eq{\label{eq:renyi_t>v}
 S^{(n)}_{[u,v]}(t>v;\beta_l,\beta_r)=\frac{1}{2}S^{(n),eq}_{[u,v]}(\beta_l)+\frac{1}{2}S^{(n),eq}_{[u,v]}(\beta_r).
}
 
We may now use similar principles in order to study the negativity.

{\ }\\
{\bf Remark.} {\em In \eqref{eqqq}, \eqref{ttch} and \eqref{split}, we expect factorization to occur independently of the pairing data of the CFT model, because only the identity module is involved in the two-point functions evaluated. Further, we expect the re-writing \eqref{rewr} to be in agreement with the original pairing between holomorphic and anti-holomorphic components imposed by the full twist fields in the original expression. This is because of the simplifications arising from taking identity modules when evaluating the limit $\varepsilon\to0$, and when evaluating the two-point function on the left subsystem.
}


\section{Evolution of the entanglement negativity after a local quench in the presence of an energy current}
\label{sec:evolution}

\begin{figure}[h!]
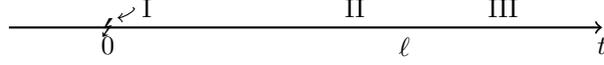

\centering\tikz[scale=1.3,anchor=base,baseline]{
\draw[thick,->]	(-1,0)--(0,0)node[anchor=north]{$0$}--(2.5,0)node[anchor=south]{II}--(3,0)node[anchor=north]{$\ell$}--(4,0)node[anchor=south]{III}--(5,0)node[anchor=north]{$t$};
\fill
	(0,-0.1)node{\Lightning};
\draw[thin,->]	(0.25,0.2)node[anchor=west]{I}to[in=0,out=0](0.1,0.1);
}
\caption{We will compute the logarithmic negativity in three regimes. I: just after the quench $t\ll 1$, II: a long time after the quench, but before the steady regime $1\ll t<\ell$ and III: in the steady state $t>\ell$.}
\label{fig:time-regimes}
\end{figure}

In the following, we will calculate the logarithmic negativity between two parts of equal length: $A_1=[-\ell,0]$ and $A_2=[0,\ell]$.
We will be considering three important time regimes: first, the time just after the quench (regime I in Figure~\ref{fig:time-regimes}), in which the numerical results in \cite{EislerZimboras2014b} suggest the logarithmic negativity grows with time logarithmically. Next we consider the system in the limit $\ell\gg t\rightarrow\infty$, which would correspond to regime II in Figure~\ref{fig:time-regimes}. From the numerical results in \cite{EislerZimboras2014b} we expect the logarithmic negativity to saturate to a constant value in this limit. Finally, when considering $\ell$ finite, one can study the NESS regime (regime III in Figure~\ref{fig:time-regimes}), which actually already exists for all times $t>\ell$. From the numerics in \cite{EislerZimboras2014b}, we expect that the value of the logarithmic negativity in this regime will again be constant in time, and lower than the value in regime II.

Using the replica trick \eqref{eq:logneg}, the logarithmic negativity at the time of the quench can be found by calculating the following expression: 
\eq{\label{eq:tr-rho-T2-t0}
\text{Tr}\rho_{[-\ell,0],[0,\ell]}^n(t=0)
\propto\langle\twist(-\ell,0)\atwist^2(0,0)\twist(\ell,0)\rangle_0,
}
where we have used the notation $\rho_{A_1,A_2}:=\rho_{A}^{T_2}$.

\begin{figure}[h!]
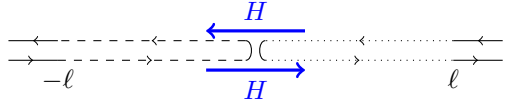

\centering\tikz[scale=1.3,anchor=base,baseline]{
\begin{scope}[decoration={
    markings,
    mark=at position 0.5 with {\arrow{>}}}]
\draw[thin,dotted,postaction={decorate}]    (2,0.2)--(0.1,0.2);
\draw[thin,dotted,postaction={decorate}]    (0.1,0)
--(2,0)node[anchor=north]{$\ell$};
\draw[thin,dashed,postaction={decorate}]    (-2,0)node[anchor=north]{$-\ell$}--(-0.1,0);
\draw[thin,dashed,postaction={decorate}]    (-0.1,0.2)--(-2,0.2);
\draw[thin,postaction={decorate}]			(-2,0.2)--(-2.5,0.2);
\draw[thin,postaction={decorate}]			(-2.5,0)--(-2,0);
\draw[thin, postaction={decorate}] 			(2,0)--(2.5,0);
\draw[thin,postaction={decorate}]			(2.5,0.2)--(2,0.2);
\draw[thin]
	(0.1,0.2)to[in=180,out=180](0.1,0)
    (-0.1,0)to[in=0,out=0](-0.1,0.2);
\draw[very thick,color=blue,->]	(0.5,0.3)--(0,0.3)node[anchor=south]{$H$}--(-0.5,0.3);
\draw[very thick,color=blue,->]	(-0.5,-0.1)--(0,-0.1)node[anchor=north]{$H$}--(0.5,-0.1);
\end{scope}
}
\caption{The negativity between two finite parts of equal length $\ell$ in an infinite system at the time of the quench $t=0$. The dashed(dotted) lines indicate that for that chiral sector, each sheet is connected with the sheet above(below). After evolution with the connected Hamiltonian $H$ the twist fields are moved into the other system, and we must regularize the expression for the negativity. 
}
\end{figure}

Since the expectation value $\langle\ldots\rangle_0$ is taken at the time of connection, we have to take into account that as the boundary condition changes at the connection point, we must regularize the expression. This will introduce various nonuniversal terms. Therefore, we will first calculate the following expression:
\eq{
\text{Tr}\rho_{A_1,A_2}^n(t=0)=c_n^2\delta^{4d_n}\langle\twist(-v,0)\atwist(-u,0)\atwist(u,0)\twist(v,0)\rangle_0,
}
where here we have defined $A_1=[-v,-u]$ and $A_2=[u,v]$.

\begin{figure}[h!]
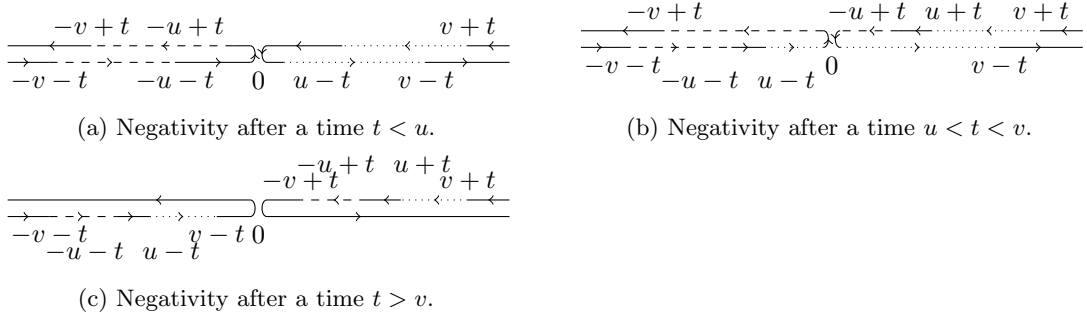

\begin{subfigure}[b]{0.5\textwidth}
\centering\tikz[scale=1.1,anchor=base,baseline]{
\begin{scope}[decoration={
    markings,
    mark=at position 0.5 with {\arrow{>}}}]
\draw[thin,dotted,postaction={decorate}]    (2.5,0.2)node[anchor=south]{$v+t$}--(1,0.2);
\draw[thin,postaction={decorate}]			(1,0.2)--(0.1,0.2)
											(0.1,0)--(0.3,0);
\draw[thin,dotted,postaction={decorate}]    (0.3,0)node[anchor=north west]{$u-t$}--(2,0)node[anchor=north]{$v-t$};
\draw[thin,dashed,postaction={decorate}]    (-2.5,0)node[anchor=north]{$-v-t$}--(-1,0)node[anchor=north]{$-u-t$};
\draw[thin,postaction={decorate}]			(-1,0)--(-0.1,0)
											(-0.1,0.2)--(-0.3,0.2);
\draw[thin,dashed,postaction={decorate}]    (-0.3,0.2)node[anchor=south east]{$-u+t$}--(-2,0.2)node[anchor=south]{$-v+t$};
\draw[thin,postaction={decorate}]			(-2,0.2)--(-3,0.2);
\draw[thin,postaction={decorate}]			(-3,0)--(-2.5,0);
\draw[thin,postaction={decorate}]			(-0.1,0)to[in=0,out=0](-0.1,0.2);
\draw[thin, postaction={decorate}] 			(2,0)--(3,0);
\draw[thin,postaction={decorate}]			(3,0.2)--(2.5,0.2);
\draw[thin,postaction={decorate}]			(0.1,0.2)to[in=180,out=180](0.1,0);
\fill
	(0,0)node[anchor=north]{$0$};
\end{scope}
}
\caption{Negativity after a time $t<u$.
}
\label{subfig:negativity-t<u}
\end{subfigure}
\begin{subfigure}[b]{0.5\textwidth}
\centering\tikz[scale=1.1,anchor=base,baseline]{
\begin{scope}[decoration={
    markings,
    mark=at position 0.5 with {\arrow{>}}}]
\draw[thin,dotted,postaction={decorate}]    (2.5,0.2)node[anchor=south]{$v+t$}--(1.2,0.2);
\draw[thin,postaction={decorate}]			(1.2,0.2)--(0.8,0.2);
\draw[thin,dashed,postaction={decorate}]	(0.8,0.2)--(0.1,0.2);
\draw[thin,dotted,postaction={decorate}]    (0.1,0)--(2,0)node[anchor=north]{$v-t$};
\draw[thin,dashed,postaction={decorate}]    (-2.5,0)node[anchor=north]{$-v-t$}--(-1.2,0);
\draw[thin,postaction={decorate}]			(-1.2,0)--(-0.8,0);
\draw[thin,dotted,postaction={decorate}]	(-0.8,0)--(-0.1,0);
\draw[thin,dashed,postaction={decorate}]    (-0.1,0.2)--(-2,0.2)node[anchor=south]{$-v+t$};
\draw[thin,postaction={decorate}]			(-2,0.2)--(-3,0.2);
\draw[thin,postaction={decorate}]			(-3,0)--(-2.5,0);
\draw[thin,postaction={decorate}]			(-0.1,0)to[in=0,out=0](-0.1,0.2);
\draw[thin, postaction={decorate}] 			(2,0)--(3,0);
\draw[thin,postaction={decorate}]			(3,0.2)--(2.5,0.2);
\draw[thin,postaction={decorate}]			(0.1,0.2)to[in=180,out=180](0.1,0);
\fill
	(0,0)node[anchor=north]{$0$}
    (-1,-0.18)node[anchor=north east]{$-u-t$}
    (-1,-0.18)node[anchor=north west]{$u-t$}
    (1,0.2)node[anchor=south east]{$-u+t$}
    (1,0.2)node[anchor=south west]{$u+t$};
\end{scope}
}
\caption{Negativity after a time $u<t<v$.
}
\label{subfig:negativity-u<t<v}
\end{subfigure}
\begin{subfigure}[b]{0.5\textwidth}
\centering\tikz[scale=1.1,anchor=base,baseline]{
\begin{scope}[decoration={
    markings,
    mark=at position 0.5 with {\arrow{>}}}]
\draw[thin,dotted,postaction={decorate}]    (2.5,0.2)node[anchor=south]{$v+t$}--(1.7,0.2);
\draw[thin,dashed,postaction={decorate}]	(1.3,0.2)--(0.5,0.2)node[anchor=south]{$-v+t$};
\draw[thin,dashed,postaction={decorate}]    (-2.5,0)node[anchor=north]{$-v-t$}--(-1.7,0);
\draw[thin,dotted,postaction={decorate}]	(-1.3,0)--(-0.5,0)node[anchor=north]{$v-t$};
\draw[thin,postaction={decorate}]			(-3,0)--(-2.5,0);
\draw[thin,postaction={decorate}]			(-0.5,0)--(-0.1,0)to[in=0,out=0](-0.1,0.2)--(-3,0.2);
\draw[thin,postaction={decorate}]			(3,0.2)--(2.5,0.2);
\draw[thin,postaction={decorate}]			(0.5,0.2)--(0.1,0.2)to[in=180,out=180](0.1,0)--(3,0);
\draw[thin,postaction={decorate}]			(-1.7,0)--(-1.3,0);
\draw[thin,postaction={decorate}]			(1.7,0.2)--(1.3,0.2);
\fill
	(0,0)node[anchor=north]{$0$}
    (-1.5,-0.2)node[anchor=north east]{$-u-t$}
    (-1.5,-0.2)node[anchor=north west]{$u-t$}
    (1.5,0.4)node[anchor=south east]{$-u+t$}
    (1.5,0.4)node[anchor=south west]{$u+t$};
\end{scope}
}
\caption{Negativity after a time $t>v$.
}
\label{subfig:negativity-t>v}
\end{subfigure}
\caption{The negativity at time $t$ between two finite parts of length $|u-v|$ evolved back to the time of the quench. On the top left (Figure~\ref{subfig:negativity-t<u}), we have the regime $t<u$, in which there is still no entanglement. The top right picture (Figure~\ref{subfig:negativity-u<t<v}) represents the case in which two cuts cross the defect. 
In the bottom picture (Figure~\ref{subfig:negativity-t>v}), the cuts have moved into the different systems, and we are in the steady regime (note that for $t>v$, the negativity does not depend on $t$).
}
\label{fig:negativity_chiral_evolution_u_v}
\end{figure}

As for the case of the EE discussed in section~\ref{sec:twistfields}, the regularization may change if the expression \eqref{eq:tr-rho-T2-t0} is evolved over time.
First, in the trivial case of $t<u$, and for the moment assuming trivial pairing data, we have:
\begin{multline}
\text{Tr}\rho_{A_1,A_2}^n(t<u;\beta_l,\beta_r)=c_n^2\delta^{4d_n}\langle\twist^{+}(-v-t,0)\atwist^{+}(-u-t,0)\atwist^{-}(-u+t,0)\twist^{-}(-v+t,0)\rangle_{\lt}\\
\langle\twist^{-}(v+t,0)\atwist^{-}(u+t,0)\atwist^{+}(u-t,0)\twist^{+}(v-t,0)\rangle_{\rt}.
\end{multline}
This can be mapped to holomorphic twist fields using \eqref{eq:unfoldingmap}, to give
\begin{multline}
\text{Tr}\rho_{A_1,A_2}^n(t<u;\beta_l,\beta_r)=c_n^2\delta^{4d_n}\langle\chtwist(-v-t)\chatwist(-u-t)\chtwist(u-t)\chatwist(v-t)\rangle^{ch}_{\beta_l}\\
\langle\chatwist(-v-t)\chtwist(-u-t)\chatwist(u-t)\chtwist(v-t)\rangle^{ch}_{\beta_r}.
\end{multline}
Using translation invariance of the holomorphic correlators, it is clear that this expression is independent of time. What's more, the correlators on the right-hand side can be rewritten as follows
\eq{
c_n\delta^{2d_n}\langle\chtwist(-v)\chatwist(-u)\chtwist(u)\chatwist(v)\rangle^{ch}_{\beta}
=:\left(\text{Tr}(\rho_{A_1\cup A_2,\emptyset}^{eq})^{n}(\beta)\right)^{1/2},
}
resulting in the following relation for the logarithmic negativity
\eq{\label{eq:logneg_t<u}
\logneg_{A_1,A_2}(t<u;\beta_l,\beta_r)=\frac{1}{2}\left(\logneg^{eq}_{A_1\cup A_2,\emptyset}(\beta_l)+\logneg^{eq}_{A_1\cup A_2,\emptyset}(\beta_r)\right)=0.
}
This just tells us what we already know: if you consider two intervals a distance $u$ away from the point of connection, at a time $t<u$ after connection, the intervals have not yet had time to build up entanglement.
The terms $\logneg^{eq}_{\tilde{A}_1,\tilde{A}_2}(\beta)$ denote the logarithmic negativity between subsystems $\tilde{A}_1$ and $\tilde{A}_2$ for a system in equilibrium at inverse temperature $\beta$. These are calculated in an infinite system where no quench has taken place. The upshot is that we can obtain time dependent results using equilibrium (finite temperature) expressions. However, as the intervals $\tilde{A}_1$ and $\tilde{A}_2$ change during the time evolution, the correlation functions may become more complicated.

If the expression \eqref{eq:tr-rho-T2-t0} is evolved over a time $u<t<v$, extra fields must be inserted at positions $(-t-\varepsilon,t)$, $(-t+\varepsilon,t)$, $(t-\varepsilon,t)$ and $(t+\varepsilon,t)$. For that, we use the following identity:
\begin{multline}
\langle\twist(-v,t)\atwist(-u,t)\atwist(u,t)\twist(v,t)\rangle_0\\
=\lim_{\varepsilon\rightarrow 0}(2\varepsilon)^{4d_n}\langle\twist(-v,t)\atwist(-t-\varepsilon,t)\twist(-t+\varepsilon,t)\atwist(-u,t)\atwist(u,t)\twist(t-\varepsilon,t)\atwist(t+\varepsilon,t)\twist(v,t)\rangle_0.
\end{multline}
With this, we can express the trace using the chiral twist fields of section~\ref{subsec:chiral-twistfields}, again assuming the CFT model in question has trivial pairing data:
\begin{multline}
\text{Tr}\rho_{A_1,A_2}^n(u<t<v;\beta_l,\beta_r)=c_n^2\delta^{8\cdim}\lim_{\varepsilon\rightarrow 0}(2\varepsilon)^{8\cdim}\\
\langle\twist^{+}(-v-t,0)\twist^{-}(-v+t,0)\atwist^{+}(-2t-\varepsilon,0)\atwist^{-}(-\varepsilon,0)\twist^{+}(-2t+\varepsilon,0)\atwist^{+}(-u-t,0)\atwist^{+}(u-t,0)\twist^{+}(-\varepsilon,0)\rangle_{\lt}\\
\langle\twist^{-}(\varepsilon,0)\atwist^{-}(-u+t,0)\atwist^{-}(u+t,0)\twist^{-}(2t-\varepsilon,0)\atwist^{-}(2t+\varepsilon,0)\atwist^{+}(\varepsilon,0)\twist^{-}(v+t,0)\twist^{+}(v-t,0)\rangle_{\rt}.
\end{multline}
Using the OPEs \eqref{eq:ope-t-at}, we can simplify this expression:
\begin{multline}
\text{Tr}\rho_{A_1,A_2}^n(u<t<v;\beta_l,\beta_r)=c_n^2\delta^{8\cdim}\lim_{\varepsilon\rightarrow 0}(2\varepsilon)^{4\cdim}\\
\langle\twist^{+}(-v-t,0)\twist^{-}(-v+t,0)\atwist^{-}(-\varepsilon,0)\atwist^{+}(-u-t,0)\atwist^{+}(u-t,0)\twist^{+}(-\varepsilon,0)\rangle_{\lt}\\
\langle\twist^{-}(\varepsilon,0)\atwist^{-}(-u+t,0)\atwist^{-}(u+t,0)\atwist^{+}(\varepsilon,0)\twist^{-}(v+t,0)\twist^{+}(v-t,0)\rangle_{\rt}.
\end{multline}
After mapping this expression to an expression containing holomorphic twist fields and using translation invariance to shift by $t$, we get
\begin{multline}
\text{Tr}\rho_{A_1,A_2}^n(u<t<v;\beta_l,\beta_r)=c_n^2\delta^{8\cdim}\lim_{\varepsilon\rightarrow 0}(2\varepsilon)^{4\cdim}
\langle\chtwist(-v)\chatwist(-u)\chatwist(u)\chtwist(t-\varepsilon)\chtwist(t+\varepsilon)\chatwist(v)\rangle^{ch}_{\beta_l}\\
\langle\chatwist(-v)\chtwist(-u)\chtwist(u)\chatwist(t-\varepsilon)\chatwist(t+\varepsilon)\chtwist(v)\rangle^{ch}_{\beta_r}.
\end{multline}
Using the OPE $\chtwist(x)\chtwist(y)\sim C_{\tau\tau}^{\tau^2}(x-y)^{\cdimsq-2\cdim}\chtwist^2(y)$, we can write this as
\begin{multline}
\text{Tr}\rho_{A_1,A_2}^n(u<t<v;\beta_l,\beta_r)=c_n^2\delta^{8\cdim}\lim_{\varepsilon\rightarrow 0}(2\varepsilon)^{2\cdimsq}(C_{\tau\tau}^{\tau^2})^2
\langle\chtwist(-v)\chatwist(-u)\chatwist(u)\chtwist^2(t)\chatwist(v)\rangle^{ch}_{\beta_l}\\
\langle\chatwist(-v)\chtwist(-u)\chtwist(u)\chatwist^2(t)\chtwist(v)\rangle^{ch}_{\beta_r}.
\end{multline}
Setting $2\varepsilon$ proportional to the cutoff parameter $\delta$ introduces an $n$-dependent constant $b_n$, defined in \eqref{eq:b_n}. This gives:
\begin{multline}
\text{Tr}\rho_{A_1,A_2}^n(u<t<v;\beta_l,\beta_r)=b_n^{2\cdimsq}c_n^2\delta^{8\cdim+2\cdimsq}(C_{\tau\tau}^{\tau^2})^2
\langle\chtwist(-v)\chatwist(-u)\chatwist(u)\chtwist^2(t)\chatwist(v)\rangle^{ch}_{\beta_l}\\
\langle\chatwist(-v)\chtwist(-u)\chtwist(u)\chatwist^2(t)\chtwist(v)\rangle^{ch}_{\beta_r}.
\end{multline}
We observe that
\eq{
c_n(c^{(2)}_n)^{1/4}\delta^{4\cdim+\cdimsq}\langle\chtwist(-v)\chatwist(-u)\chatwist(u)\chtwist^2(t)\chatwist(v)\rangle^{ch}_{\beta}=:\left(\text{Tr}\rho_{\tilde{A}_1,\tilde{A}_2}^{n,eq}\right)^{1/2},
}
where $\tilde{A}_1=[-v,-u]\cup[t,v]$, and $\tilde{A}_2=[u,t]$.

Taking the log of this expression, and sending $n\rightarrow 1$ from $n$ even in \eqref{eq:cdimsq},
\eq{
\lim_{\substack{n\rightarrow 1 \\ n\text{ even}}}\cdimsq = -\frac{c}{8},
}
we can express the logarithmic negativity a time $u<t<v$ after the quench for CFTs with trivial pairing in terms of the logarithmic negativity of systems in equilibrium at temperatures $\beta_l$ and $\beta_r$, respectively:
\eq{\label{eq:logneg_u<t<v}
\logneg_{A_1,A_2}(u<t<v;\beta_l,\beta_r)=\frac{1}{2}\logneg^{eq}_{\tilde{A}_1,\tilde{A}_2}(\beta_l)+\frac{1}{2}\logneg^{eq}_{\tilde{A}_1,\tilde{A}_2}(\beta_r)+\ln\Ctttsq -\ln c_{1/2}-\frac{c}{4}\ln b_1,
}
where we have used $\Ctttsq=(C_{\tau\tau}^{\tau^2})^2$. Note that the structure constants $\Ctttsq$ defined in the OPE depend on $n$, and that in \eqref{eq:logneg_u<t<v} the limit $n\rightarrow 1$ from even $n$ has been taken. We will not use separate notation to indicate this. Also note that in \eqref{eq:logneg_u<t<v} we used $c^{(2)}_n=c_{n/2}^2$ for $n$ even. The last term, $-\frac{c}{4}\ln b_1=3\ln g$, which is just three times the boundary entropy (see Appendix~\ref{sec:boundary_entropy}). 
Again, the terms $\logneg^{eq}_{\tilde{A}_1,\tilde{A}_2}(\beta)$ denote the logarithmic negativity between subsystems $\tilde{A}_1$ and $\tilde{A}_2$ for a system in equilibrium at inverse temperature $\beta$. As before, the effect of the quench is encoded in the now changed intervals $\tilde{A}_1=[-v,-u]\cup[t,v]$ and $\tilde{A}_2=[u,t]$.


The expression for $\text{Tr}\rho_{A_1,A_2}^n(t)$ at late times $t>v$ does not need a regulator, as in that case the cuts do not extend over the connection point.
We have
\begin{multline}
\text{Tr}\rho_{A_1,A_2}^n(t>v;\beta_l,\beta_r)=c_n^2\delta^{4d_n}\langle\twist^{+}(-v-t,0)\atwist^{+}(-u-t,0)\atwist^{+}(u-t,0)\twist^{+}(v-t,0)\rangle_{\lt}\\
\langle\twist^{-}(-v+t,0)\atwist^{-}(-u+t,0)\atwist^{-}(u+t,0)\twist^{-}(v+t,0)\rangle_{\rt},
\end{multline}
which, after mapping to holomorphic twist fields and using translation invariance, becomes
\eq{
\text{Tr}\rho_{A_1,A_2}^n(t>v;\beta_l,\beta_r)=c_n^2\delta^{4d_n}\langle\chtwist(-v)\chatwist(-u)\chatwist(u)\chtwist(v)\rangle^{ch}_{\beta_l}\langle\chatwist(-v)\chtwist(-u)\chtwist(u)\chatwist(v)\rangle^{ch}_{\beta_r}.
}
This results in the following expression
\eq{\label{eq:logneg_t>v}
\logneg_{A_1,A_2}(t>v;\beta_l,\beta_r)=\frac{1}{2}\logneg^{eq}_{A_1,A_2}(\beta_l)+\frac{1}{2}\logneg^{eq}_{A_1,A_2}(\beta_r).
}

From the form of \eqref{eq:logneg_t<u}, \eqref{eq:logneg_u<t<v} and \eqref{eq:logneg_t>v}, we can counclude that for intervals $A_1=[-v,-u]$ and $A_2=[u,v]$ that are equidistant from the connection point and have equal length, we can always write the logarithmic negativity after the quench as the average of the logarithmic negativity for a system that is thermalized at inverse temperature $\beta_l$ and one that is thermalized at temperature $\beta_r$:
\eq{\label{eq:logneg-equal-lengths}
\logneg_{[-v,-u],[u,v]}(t;\beta_l,\beta_r)=\frac{1}{2}\left(\logneg_{[-v,-u],[u,v]}(t;\beta_l)+\logneg_{[-v,-u],[u,v]}(t;\beta_r)\right).
}
Note that for models with trivial pairing data, this expression is valid at any time $t$ after the quench. However, for more general CFTs we expect this to hold only for $t>v$. In the following, we will always calculate the negativity for the case $\beta_l=\beta_r=\beta$. It must therefore be noted that all our results for $t<v$ may have corrections.

Let us now specialize to the case $A_1=[-\ell,0]$ and $A_2=[0,\ell]$. In this case there are two situations: $t<\ell$ and $t>\ell$. We must substitute $v=\ell$ and take the limit $u\rightarrow 0$ in the expressions \eqref{eq:logneg_u<t<v} and \eqref{eq:logneg_t>v}.

\begin{figure}[h!]
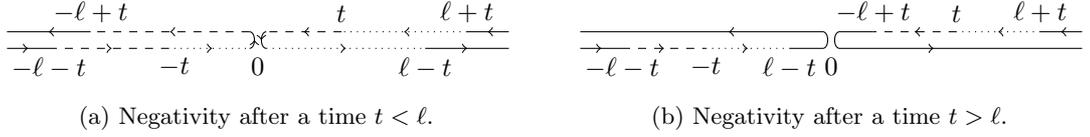

\begin{subfigure}[b]{0.5\textwidth}
\centering\tikz[scale=1.1,anchor=base,baseline]{
\begin{scope}[decoration={
    markings,
    mark=at position 0.5 with {\arrow{>}}}]
\draw[thin,dotted,postaction={decorate}]    (2.5,0.2)node[anchor=south]{$\ell+t$}--(1,0.2)node[anchor=south]{$t$};
\draw[thin,dashed,postaction={decorate}]	(1,0.2)--(0.1,0.2);
\draw[thin,dotted,postaction={decorate}]    (0.1,0)--(2,0)node[anchor=north]{$\ell-t$};
\draw[thin,dashed,postaction={decorate}]    (-2.5,0)node[anchor=north]{$-\ell-t$}--(-1,0)node[anchor=north]{$-t$};
\draw[thin,dotted,postaction={decorate}]	(-1,0)--(-0.1,0);
\draw[thin,dashed,postaction={decorate}]    (-0.1,0.2)--(-2,0.2)node[anchor=south]{$-\ell+t$};
\draw[thin,postaction={decorate}]			(-2,0.2)--(-3,0.2);
\draw[thin,postaction={decorate}]			(-3,0)--(-2.5,0);
\draw[thin,postaction={decorate}]			(-0.1,0)to[in=0,out=0](-0.1,0.2);
\draw[thin, postaction={decorate}] 			(2,0)--(3,0);
\draw[thin,postaction={decorate}]			(3,0.2)--(2.5,0.2);
\draw[thin,postaction={decorate}]			(0.1,0.2)to[in=180,out=180](0.1,0);
\fill
	(0,0)node[anchor=north]{$0$};
\end{scope}
}
\caption{Negativity after a time $t<\ell$.
}
\label{subfig:negativity-small_t}
\end{subfigure}
\begin{subfigure}[b]{0.5\textwidth}
\centering\tikz[scale=1.1,anchor=base,baseline]{
\begin{scope}[decoration={
    markings,
    mark=at position 0.5 with {\arrow{>}}}]
\draw[thin,dotted,postaction={decorate}]    (2.5,0.2)node[anchor=south]{$\ell+t$}--(1.5,0.2)node[anchor=south]{$t$};
\draw[thin,dashed,postaction={decorate}]	(1.5,0.2)--(0.5,0.2)node[anchor=south]{$-\ell+t$};
\draw[thin,dashed,postaction={decorate}]    (-2.5,0)node[anchor=north]{$-\ell-t$}--(-1.5,0)node[anchor=north]{$-t$};
\draw[thin,dotted,postaction={decorate}]	(-1.5,0)--(-0.5,0)node[anchor=north]{$\ell-t$};
\draw[thin,postaction={decorate}]			(-3,0)--(-2.5,0);
\draw[thin,postaction={decorate}]			(-0.5,0)--(-0.1,0)to[in=0,out=0](-0.1,0.2)--(-3,0.2);
\draw[thin,postaction={decorate}]			(3,0.2)--(2.5,0.2);
\draw[thin,postaction={decorate}]			(0.5,0.2)--(0.1,0.2)to[in=180,out=180](0.1,0)--(3,0);
\fill
	(0,0)node[anchor=north]{$0$};
\end{scope}
}
\caption{Negativity after a time $t>\ell$.
}
\label{subfig:negativity-large_t}
\end{subfigure}
\caption{The negativity at time $t$ between two finite parts of length $\ell$ evolved back to the time of the quench. On the left (Figure~\ref{subfig:negativity-small_t}), we have the case $t<\ell$, in which two cuts cross the defect and when expressing the negativity in the disconnected system, the 3-point functions become 4-point functions. On the right (Figure~\ref{subfig:negativity-large_t}), the cuts have moved into the different systems, and we are in the steady regime (note that for $t>\ell$, the negativity does not depend on $t$).
}
\label{fig:negativity_chiral_evolution}
\end{figure}

In order to find the relation between the logarithmic negativity a time $t$ after the quench and the negativity in equilibrium systems, we take the appropriate limits from the physical quantities we have computed. For instance, the equilibrium expressions are defined as
\eq{\label{eq:logneg_lim_uv_0l}
\logneg^{eq}_{[-\ell,0],[0,\ell]}(\beta):=\lim_{\substack{u\rightarrow 0 \\ v\rightarrow \ell}}\logneg^{eq}_{[-v,-u],[u,v]}(\beta).
}
Using \eqref{eq:logneg_lim_uv_0l}, \eqref{eq:logneg_u<t<v} and \eqref{eq:logneg_t>v}, we deduce that
\eq{
\begin{split}
\logneg_{[-\ell,0],[0,\ell]}(t;\beta)
&=\lim_{\substack{u\rightarrow 0 \\ v\rightarrow \ell}}\logneg_{[-v,-u],[u,v]}(t;\beta).
\end{split}
}
As a check, we may compute the relation between the two choices of intervals explicitly, and we find that irrespective of the state of the system, we get the following expression:
\eq{
\begin{split}
\lim_{\substack{u\rightarrow 0 \\ v\rightarrow \ell}}\text{Tr}(\rho_{[-v,-u],[u,v]})^n&=\lim_{\substack{u\rightarrow 0 \\ v\rightarrow \ell}}c_n^2\delta^{4d_n}\langle\twist(-v)\atwist(-u)\atwist(u)\twist(v)\rangle\\
&=c_n^2C_{\atwist\atwist}^{\atwist^2}a_n^{-2d_n+d_n^{(2)}}\delta^{2d_n+d_n^{(2)}}\langle\twist(-\ell)\atwist^2(0)\twist(\ell)\rangle,
\end{split}
}
where in the second step we used the OPE $\atwist(x)\atwist(y)\sim C_{\atwist\atwist}^{\atwist^2}(x-y)^{-2d_n+d_n^{(2)}}\atwist^2(y)$, 
and the constant $a_1$ appears when we set $(x-y)\sim a_n\delta$ in this OPE. This constant is different from $b_n$, since the OPE is different (it gives a different change in geometry).
We note that for both equilibrium and nonequilibrium states of the system we encounter the same combination of constants, which on physical grounds we require to add up to zero:
\eq{
\frac{c}{4}\ln a_1+\ln c_{1/2}-\ln C_{\atwist\atwist}^{\atwist^2}=0.
}


\subsection{Early times: regimes I and II ($t<\ell$)}
\label{subsec:regime_I_II}

The expressions $\logneg^{eq}_{[-\ell,0]\cup[t,\ell],[0,t]}(\beta)$ contain four-point functions:
\eq{
\text{Tr}(\rho^{eq}_{[-\ell,0]\cup[t,\ell],[0,t]})^n(\beta)=c_nc^{(2)}_n\delta^{2d_n+2d^{(2)}_n}\langle\twist(-\ell)\atwist^2(0)\twist^2(t)\atwist(\ell)\rangle_{\beta}.
}
In this case the general result is strongly model dependent, as the expression for a four-point function contains a model-dependent function of the cross-ratio of the four coordinates. We first map this correlation function to the plane
\eq{
\langle\twist(-\ell)\atwist^2(0)\twist^2(t)\atwist(\ell)\rangle_{\beta}=\left(\frac{2\pi}{\beta}\right)^{2d_n+2d^{(2)}_n}(e^{2\pi t/\beta})^{d^{(2)}_n}\langle\twist(e^{-2\pi\ell/\beta})\atwist^2(1)\twist^2(e^{2\pi t/\beta})\atwist(e^{2\pi\ell/\beta})\rangle_{\mathbb{C}}.
}
Using global conformal invariance, the four-point function on the plane can be brought in the following form:
\eq{\label{eq:fourpoint-squares}
\langle\twist(z_1)\atwist^2(z_2)\twist^2(z_3)\atwist(z_4)\rangle_{\mathbb{C}}=|z_{14}|^{-2d_n}|z_{23}|^{-2d^{(2)}_n}\mathcal{F}_n(\eta),
}
with the four-point ratio
\eq{
\eta=\frac{z_{12}z_{34}}{z_{13}z_{24}}.
}
Mapping this result back to the cylinder using $w_i=\frac{\beta}{2\pi}\ln z_i$, and using 
\eq{\label{eq:lim_dimensions}
\lim_{\substack{n\rightarrow 1 \\ n\text{ even}}}d_{n}=0,\qquad\text{and}\qquad\lim_{\substack{n\rightarrow 1 \\ n\text{ even}}}d^{(2)}_{n}=-c/4,
}
we can express the finite-temperature negativity between $\tilde{A}_1=[-\ell,0]\cup[t,\ell]$ and $\tilde{A}_2=[0,t]$, for $t<\ell$ as follows
\eq{\label{eq:negt<l}
\logneg^{eq}_{[-\ell,0]\cup[t,\ell],[0,t]}(\beta)=\frac{c}{2}\ln\left(\frac{\beta}{\pi\delta}\sinh\frac{\pi t}{\beta}\right)+f\left(\frac{\sinh(\pi(\ell-t)/\beta)}{\sinh(\pi(\ell+t)/\beta)}\right)+2\ln c_{1/2},\qquad t<\ell.
}
The function $f(\eta):=\lim_{\substack{n\rightarrow 1 \\ n\text{ even}}}\mathcal{F}_n(\eta)$ is model dependent (it depends on the universality class of the CFT model). However, we may find its value for general CFT in certain limits, where the four-point function reduces to a two- or three-point function.

\subsubsection{Regime I}
\label{subsubsec:regime_I}
The behaviour right after the quench can be studied directly by taking the limit $t\rightarrow 0$, which corresponds to $z_3\rightarrow z_2$ in \eqref{eq:fourpoint-squares}, and considering the OPE
\eq{
\atwist^2(x)\twist^2(y)\stackrel{x\rightarrow y}{\sim}|x-y|^{-2d^{(2)}_n}.
}
With this, the four-point function \eqref{eq:fourpoint-squares} becomes
\eq{
\langle\twist(z_1)\atwist^2(z_2)\twist^2(z_3)\atwist(z_4)\rangle
\stackrel{z_2\rightarrow z_3}{\sim}|z_{23}|^{-2d^{(2)}_n}
\langle\twist(z_1)\atwist(z_4)\rangle.
}
Again mapping back to the cylinder using $z_j:=\exp(2\pi w_j/\beta)$, with $w_1=-\ell$, $w_2=0$, $w_3=t$ and $w_4=\ell$, and considering the scaling dimensions \eqref{eq:lim_dimensions} in the limit $n\rightarrow 1$ from even values of $n$, we find that, for very early times, the behaviour of the logarithmic negativity is characterised by the function
\eq{
\logneg^{eq}_{[-\ell,0]\cup[t,\ell],[0,t]}(\beta)\sim\frac{c}{2}\ln\frac{t}{\delta} + 2\ln c_{1/2},\qquad t\ll\text{ any other scale}.
}
Note that this equation only holds for $t$ very small, but the constants in the expression are all accounted for.
Using \eqref{eq:logneg_u<t<v} to compute the logarithmic negativity after the quench, we obtain the dynamics for the negativity a very short time after the quench.
\eq{\label{eq:logneg_t->0}
\logneg_{[-\ell,0],[0,\ell]}(t;\beta)\sim\frac{c}{2}\ln\frac{t}{\delta}+\ln\Ctttsq+ \ln c_{1/2}+3\ln g
,\qquad t\ll\text{ any other scale},
}
where again we note that $-\frac{c}{4}\ln b_1=3\ln g$ (see Appendix~\ref{sec:boundary_entropy}), where $\ln g$ is the boundary entropy \cite{AffleckLudwig1991}. Note that although this result has been derived using the assumption of trivial pairing data, we expect this result to hold for any CFT.

\subsubsection{Regime II}
\label{subsubsec:regime_II}
Another regime in which we may find a general expression is the limit $\ell\gg t\rightarrow \infty$. Note that the cross ratio in \eqref{eq:negt<l}  depends on $\ell$ and $t$. After taking the limit $\ell\rightarrow\infty$, the cross ratio reduces to $\exp(-2\pi t/\beta)$, and the four-point function simplifies to
\eq{\label{eq:neg-largeL}
\lim_{\substack{n\rightarrow 1\\ n\text{ even}}}\lim_{\ell\rightarrow\infty}\langle\twist(-\ell)\atwist^2(0)\twist^2(t)\atwist(\ell)\rangle_{\beta}
=\left(\frac{\beta}{\pi}\sinh\frac{\pi t}{\beta}\right)^{c/2}\mathcal{F}_1\left(e^{-2\pi t/\beta}\right).
}
This gives model dependent behaviour of the logarithmic negativity as a function of time, since there is a time-dependent part in $f$ that is dependent on the CFT model (or specifically, its universality class), and must be computed for different models, but we may study the limiting behaviour of $f(e^{-2\pi t/\beta})$ for  very early or very late times. Note that we do not expect to reach the NESS regime at late times, since we consider $t<\ell$. 

To characterise the behaviour of $f(\eta)$ for $\eta\rightarrow 0$, we can compare the general expression for the four-point function, and take the limit $z_1\rightarrow z_2$, so that the cross ratio $\eta\rightarrow 0$. On the other hand, we may evaluate the lhs explicitly by using the OPE
\eq{
\twist(z_1)\atwist^2(z_2)\stackrel{z_1\rightarrow z_2}{\sim}\Ctatsqat\,|z_{12}|^{-d^{(2)}_n}\atwist(z_2).
}
By inserting an extra twist field, and comparing the expectation value of the lhs as $z_1\rightarrow z_2$ to the expectation value of the rhs, we obtain the following relation for the structure constant $\Ctatsqat$:
\eq{\label{eq:relation_str-cst_threepoint}
\Ctatsqat=\lim_{z_1\rightarrow z_2}\Ctatsqt\;|z_{13}|^{-d^{(2)}_n}|z_{23}|^{d^{(2)}_n}=\Ctatsqt.
}
Using this OPE, we can calculate the four-point function in the limit $z_1\rightarrow z_2$.
\eq{
\begin{split}
\langle\twist(z_1)\atwist^2(z_2)\twist^2(z_3)\atwist(z_4)\rangle
&\stackrel{z_1\rightarrow z_2}{\sim}\Ctatsqat\;|z_{12}|^{-d^{(2)}_n}\langle\atwist(z_2)\twist^2(z_3)\atwist(z_4)\rangle\\
&=(\Ctatsqat)^2\;|\eta|^{-d^{(2)}_n}|z_{23}|^{-2d^{(2)}_n}|z_{24}|^{-2d_n},
\end{split}
}
where we used \eqref{eq:relation_str-cst_threepoint} to obtain $C_{\atwist\twist^2\atwist}=\Ctatsqt=\Ctatsqat$.
Comparing this with \eqref{eq:fourpoint-squares}, we see that the function $\mathcal{F}_n$ behaves in the limit $\eta\rightarrow 0$ as
\eq{
\mathcal{F}_n(\eta)\stackrel{\eta\rightarrow 0}{\sim} (\Ctatsqat)^2\;|\eta|^{-d^{(2)}_n}.
}
The result for the negativity in the limit $\ell\gg t\rightarrow\infty$ is
\eq{
\lim_{\substack{n\rightarrow 1 \\ n\text{ even}}}\lim_{\ell\gg t\rightarrow\infty}\langle\twist(-\ell)\atwist^2(0)\twist^2(t)\atwist(\ell)\rangle_{\beta}=(\Ctatsqat)^2\left(\frac{\beta}{2\pi}\right)^{c/2},
}
resulting in the following expression for the equilibrium negativity for the changed interval\footnote{The expressions for the negativity in equilibrium \eqref{eq:logneg_eq_-inf0tinf_0t} for $\ell\rightarrow\infty$, correspond to the negativity of a bipartite system at finite temperature, which was calculated in \cite{CalabreseCardyTonni2014}. Our results agree, but we have made a different choice of function $\mathcal{F}_n(x)$}:
\eq{\label{eq:logneg_eq_-inf0tinf_0t}
\logneg^{eq}_{[-\infty,0]\cup[t,\infty],[0,t]}(\beta)=\frac{c}{2}\ln\frac{\beta}{2\pi\delta}+2\ln \Ctatsqat+2\ln c_{1/2},\qquad t\rightarrow\infty.
}
Finally, we substitute the above expression into \eqref{eq:logneg_u<t<v} to find the logarithmic negativity in this limit, valid for CFT models with trivial pairing data:
\begin{equation}\label{eq:logneg_t->inf}
\lim_{s\rightarrow\infty}\logneg_{[-\infty,0],[0,\infty]}(s;\beta)=\frac{c}{2}\ln\frac{\beta}{2\pi\delta}+2\ln\Ctatsqat+\ln\Ctttsq +\ln c_{1/2}+3\ln g.
\end{equation}
This is simplified by using the relation
\begin{equation}\label{eqCC}
\Ctatsqat=\Ctttsq
\end{equation}
proved in Appendix \ref{sec:structureconstants}.
The first thing we notice is that expression \eqref{eq:logneg_eq_-inf0tinf_0t} does not depend on $t$, thus confirming that the limit in \eqref{eq:logneg_t->inf} exists, and in regime II the logarithmic negativity reaches a plateau. Unsurprisingly, the height of these plateaus decreases at higher temperatures.
Another thing we may do is look at the difference of the logarithmic negativity in regime I and regime II, given by equations \eqref{eq:logneg_t->0} and \eqref{eq:logneg_t->inf}, the result is a universal function of $t$ and the inverse temperature $\beta$:
\eq{\label{univdiff}
\logneg_{[-\infty,0],[0,\infty]}(t;\beta)-\lim_{s\rightarrow\infty}\logneg_{[-\infty,0],[0,\infty]}(s;\beta)=\frac{c}{2}\ln\frac{2\pi t}{\beta}-2\ln\Ctatsqat \qquad t\ll\text{ any other scale}.
}



\subsection{Late times: regime III ($t>\ell$)}
\label{subsec:regime_III}
We compute the equilibrium expression for the trace:
\eq{
\text{Tr}(\rho^{eq}_{[-\ell,0],[0,\ell]})^n(\beta)=c_{n}(c^{(2)}_{n})^{1/2}\delta^{2d_n+d^{(2)}_n}\langle\twist(-\ell)\atwist^2(0)\twist(\ell)\rangle_{\beta}.
}
Using the exponential map, we get the expression in terms of a correlation function on the Riemann sphere:
\eq{
\langle\twist(-\ell)\atwist^2(0)\twist(\ell)\rangle_{\beta}=\left(\frac{\beta}{2\pi}\right)^{-2d_n-d^{(2)}_n}\langle\twist(e^{-2\pi\ell/\beta})\atwist^2(1)\twist(e^{2\pi\ell/\beta})\rangle_{\mathbb{C}}
}
Using \eqref{eq:relation_str-cst_threepoint}, we have
\eq{
\langle\twist(e^{-2\pi\ell/\beta})\atwist^2(1)\twist(e^{2\pi\ell/\beta})\rangle_{\mathbb{C}}
=\Ctatsqat\left(2\sinh\frac{2\pi\ell}{\beta}\right)^{-2d_{n}}\left(\tanh\frac{\pi\ell}{\beta}\right)^{-d^{(2)}_n},
}
from which we compute the equilibrium negativity:
\eq{
\logneg^{eq}_{[-\ell,0],[0,\ell]}(\beta)=\frac{c}{4}\ln\left(\frac{\beta}{2\pi\delta}\tanh\frac{\pi\ell}{\beta}\right)+\ln\Ctatsqat +\ln  c_{1/2}.
}
Using \eqref{eq:logneg_t>v} we find that this is equal to the NESS logarithmic negativity for $t>\ell$:
\eq{\label{eq:logneg_ness}
\logneg^{NESS}_{[-\ell,0],[0,\ell]}(\beta)=\logneg_{[-\ell,0],[0,\ell]}(t>\ell;\beta)=\frac{c}{4}\ln\left(\frac{\beta}{2\pi\delta}\tanh\frac{\pi\ell}{\beta}\right)+\ln\Ctatsqat +\ln c_{1/2}.
}
Note that this expression is independent of pairing data of the CFT, and therefore should hold for general CFT.

We can compare the values of this plateau (regime III) with the plateau in regime II by taking the limit $\ell\rightarrow\infty$ in \eqref{eq:logneg_ness}. The result is
\eq{
\lim_{s\rightarrow\infty}\logneg_{[-\infty,0],[0,\infty]}(s;\beta)-\logneg^{NESS}_{[-\infty,0],[0,\infty]}(\beta)=\frac{c}{4}\ln\frac{\beta}{2\pi\delta}+\ln\Ctatsqat +\ln\Ctttsq+3\ln g,
}
which is again simplified using \eqref{eqCC}.

{\ }\\
{\bf Remark.} {\em We expect the general relations \eqref{eq:logneg_t<u}, \eqref{eq:logneg_u<t<v} and \eqref{eq:logneg_t>v} 
to depend on the pairing data of the CFT model. However, the results of section~\ref{subsubsec:regime_I} are expected to hold in general, due simplifications arising when taking the limit $t\rightarrow 0$.
}

\section{Discussion / Conclusion}
We have found analytical expressions for the EE and the logarithmic negativity after a local, ``cut and glue''-type, quench that are valid for CFT models with trivial pairing data, as well as a few that are valid for any CFT, in certain time regimes. These expressions are in agreement with the behaviour found in \cite{EislerZimboras2014b}, in which the time evolution of the logarithmic negativity was studied numerically for the case of the harmonic chain. In particular, our initial logarithmic growth with $t$ appears to agree with the behaviour found in \cite{EislerZimboras2014b}, as does the initial saturation in regime II, which is reached when $t$ is large enough, but still smaller than $\ell$. The results in \cite{EislerZimboras2014b} also suggest the existence of a NESS shortly after the point $t>\ell$ is reached (in our exact results this is instantaneous), whose conjectured form is confirmed by our results.

We find that for the case $t<\ell$ the universal dependence on $t$ and $\ell$ of the logarithmic negativity has the same form as the logarithmic negativity in a thermal state between a region $[0,t]$ and its direct environment $[-\ell,0]\cup[t,\ell]$, which has been considered in \cite{CalabreseCardyTonni2014}.

The appearance of the term $2\ln\Ctatsqat$ in the universal difference \eqref{univdiff}, in particular of the factor 2, seems to indicate the appearance, at large times $t$ (in the limit $\ell\to\infty$), of two points around which the independent contributions to the entanglement arise. Looking at the form of the equilibrium expressions in \eqref{eq:logneg_u<t<v} in this limit, given by \eqref{eq:logneg_eq_-inf0tinf_0t}, it is clear from the intervals $\tilde{A}_1=[-\infty,0]\cup[t,\infty]$, $\tilde{A}_2=[0,t]$ that in the large $t$ limit the same result can be obtained by using a product of two three-point functions $\langle\twist\atwist^2\twist\rangle\langle\atwist\twist^2\atwist\rangle$, giving rise to the $2\ln\Ctatsqt$ term (note that $C_{\atwist\twist^2\atwist}=\Ctatsqt$). This observation on the equilibrium expression represents the fact that the negativity of an interval of length $t$ with respect to the rest of the system at finite temperatures has, at large $t$, two independent contributions coming from the boundary points of an interval, due to the finite effective correlation length generated by the nonzero temperature.

\begin{figure}[here]
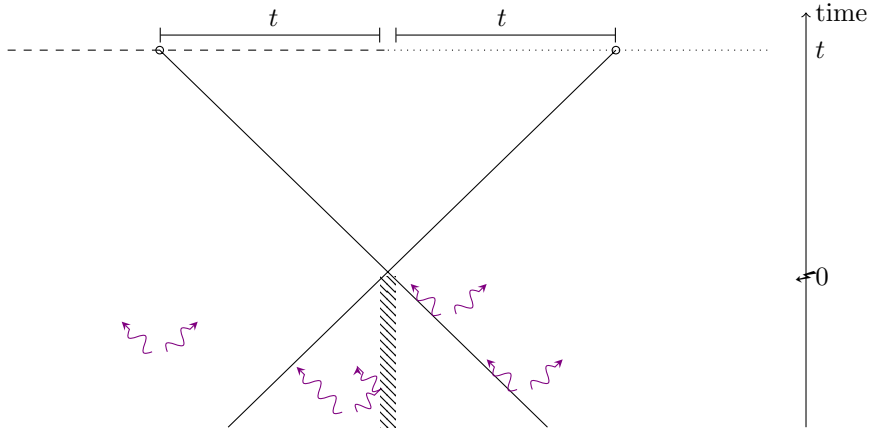

\centering\tikz{
\draw[thin,dashed]		(-5,3)--(0,3);
\draw[thin,dotted]		(0,3)--(5,3);
\draw[thin,|-|]			(-3,3.2)--(-1.5,3.2)node[anchor=south]{$t$}--(-0.1,3.2);
\draw[thin,|-|]			(0.1,3.2)--(1.5,3.2)node[anchor=south]{$t$}--(3,3.2);
\draw[thin]				(0.1+2,0-2)--(-3,3)circle(0.05)
						(-0.1-2,0-2)--(3,3)circle(0.05);
\draw[thin,snake=coil, line after snake=0.5mm, segment aspect=0,%
        segment length=8pt,color=red!50!blue,-stealth]
                        (-3.1,-1)--(-3.5,-0.6);
\draw[thin,snake=coil, line after snake=0.5mm, segment aspect=0,%
        segment length=8pt,color=red!50!blue,-stealth]
                        (-2.9,-1)--(-2.5,-0.6);
\draw[thin,snake=coil, line after snake=0.5mm, segment aspect=0,%
        segment length=8pt,color=red!50!blue,-stealth]
                        (0.7,-0.5)--(0.3,-0.1);
\draw[thin,snake=coil, line after snake=0.5mm, segment aspect=0,%
        segment length=8pt,color=red!50!blue,-stealth]
                        (0.9,-0.5)--(1.3,-0.1);
\draw[thin,snake=coil, line after snake=0.5mm, segment aspect=0,%
        segment length=8pt,color=red!50!blue,-stealth]
                        (1.7,-1.5)--(1.3,-1.1);
\draw[thin,snake=coil, line after snake=0.5mm, segment aspect=0,%
        segment length=8pt,color=red!50!blue,-stealth]
                        (1.9,-1.5)--(2.3,-1.1);
\draw[thin,snake=coil, line after snake=0.2mm, segment aspect=0,%
        segment length=6pt,color=red!50!blue,-stealth]
                        (-0.4,-1.8)--(-0.1,-1.5)--(-0.4,-1.2);
\draw[thin,snake=coil, line after snake=0.6mm, segment aspect=0,%
        segment length=8pt,color=red!50!blue,-stealth]
                        (-0.6,-1.8)--(-1.2,-1.2);
\fill[pattern = north west lines]
						(-0.1,0) rectangle (0.1,-2);
\draw[thin,->]			(5.5,-2)--(5.5,0)node[rotate=-45]{\Lightning}node[anchor=west]{$0$}--(5.5,3)node[anchor=west]{$t$}--(5.5,3.5)node[anchor=west]{time};
}
\caption{A time $t$ after the quench, we start to see an effect of two lightcones (whose size depends on the length $\ell$ of the intervals, so they can be infinite), one starting somewhere in the left system and one in the right, where in each lightcone there is an interval of length $t$ in the other half of the system, which can entangle with the rest of the interval within the lightcone. If we take these two lightcones together, we end up with something looking like an interval of length $t$ in a larger system, which accounts for the appearance of two factors of $C_{\twist\atwist^2\twist}$ at late times.}
\label{fig:lightcone}
\end{figure}

We may also give a physical explanation for the relation \eqref{eq:logneg_u<t<v} between the negativity after the quench and the equilibrium negativity. This physical explanation accounts for the equality of the universal parts: the dependence on the time, temperatures and interval lengths, up to additional non-universal constants. We take the case $u=0$ and $v=\ell\to\infty$ for simplicity. See Figure~\ref{fig:lightcone} for a pictorial representation. In this picture, one considers the creation of entangled pairs at any time before or after the quench. In the disconnected state, any particle from an entangled pair reflects at the defect. However, after connection, one of the entangled particles can move into the other subsystem. Whether this happens, depends on the time of creation, and the distance from the connection point. Using such rules, one can ``count'' the number of entangled pairs contributing to the entanglement between the left and right after a time $t$. On the other hand, one can also count the number of entangled pairs contributing to the entanglement of an interval of length $t$ at equilibrium (without defect). A moment's thought shows that these two numbers are equal.

In our calculations we have assumed that the pairing between holomorphic and antiholomorphic modules of the CFT is trivially factorized. This is not the case in general, and therefore the relations we have found between the logarithmic negativity after a quench and the logarithmic negativity in equilibrium do not hold in general. However, as explained, in certain time regimes the results are expected to become independent of pairing data. Further, it is possible that the above physically compelling particle-pair-creation picture could have more general validity.

A next step would be to learn more about the way in which pairing affects our computations. In particular, we would like to find limits in which the results are independent of pairing, and determine the 
corrections that our general relations would get for CFT models with nontrivial pairing data. Other interesting directions are generalizing these results to integrable QFTs, and to cases with nontrivial impurities after the connection (that situation has been studied in the recent work \cite{BernardDoyonViti2014}). Another avenue would be to apply the ideas developed in this and related work to other observables.

{\bf Acknowledgements.}
The authors would like to thank Pasquale Calabrese, John Cardy and Zoltan Zimboras for useful feedback.

%


\appendix

\section{The steady-state density matrix and scattering map}
\label{sec:scattering}

Consider again the quench problem as depicted in Fig. \ref{fig:physicalsituation}, where two independently thermalized halves of the system are connected to each other and let to evolve unitarily. The steady state is reached in the region around the connection point after an infinite time evolution. More precisely, the steady-state (stationary) limit is
\eq{
	\langle {\cal O} \rangle_{\text{ness}} = \lim_{t\to\infty} \lim_{L\to\infty}
	\langle e^{iHt}{\cal O}e^{-iHt} \rangle_0
}
where $L$ is the total length of the system. This limit is expected to exist for $\cal O$ being any local observable or product thereof.

In \cite{bernard2012energy, BernardDoyon2014}, the family of observables formed by the stress-energy tensor and its descendants (the ``energy sector'') was considered. In CFT, this family can of course be factorized into right-movers and left-movers. It was shown that, on this family, the result of the steady-state limit can be described by a state where right-movers and left-movers are independently thermalized. That is, if $\varphi_1^+ \varphi_2^-$ is a product of right-moving and left-moving observables in the energy sector, then it was shown that
\eq{\label{chfa}
	\langle \varphi_1^+ \varphi_2^-\rangle_{\text{ness}} =
	\langle \varphi_1\rangle_{\beta_l}
	\langle \varphi_2\rangle_{\beta_r}
}
where $\varphi_{1,2}$ are the chiral fields corresponding to $\varphi^\pm_{1,2}$. This, in turn, can be interpreted as emerging from a simple density matrix:
\eq{
	\langle {\cal O}\rangle_{\text{ness}} = \frac{{\rm Tr}\left(e^{-W}{\cal O}\right)}{{\rm Tr}\left(e^{-W}\right)}
}
where
\eq{
	W = \beta_l H_+ + \beta_r H_-
}
and $H_\pm$ represent the total right/left-moving energies. Owing to the the fact that the total energy is $H=H_++H_-$ and that the total momentum is $P = H_+-H_-$, one can further re-interpret this density matrix as the  boost of a thermal state \cite{BhaseenDoyonLucasSchalm2013}:
\eq{
	W = \beta_{\rm rest} (\cosh\theta\;H -\sinh\theta\;P)
}
where the rest-frame inverse temperature is $\beta_{\rm rest} = \sqrt{\beta_l\beta_r}$ and the boost velocity is $\tanh\theta = (\beta_r-\beta_l)/(\beta_r+\beta_l)$.

It is interesting to extend this family of observable and determine the form of the steady state on the extended family. One of course expects the steady state to be described, on extended families, in a similar manner as above, although the sharp light cone effect describing the instantaneous reach of the steady state in the energy sector \cite{bernard2012energy, BernardDoyon2014} is not expected to hold in general.

In this section we show that the above description of the steady state stays valid on the branch-point twist fields, where the right- and left-moving factors are the right- and left-moving branch-point twist fields discussed in Section \ref{sec:twistfields}.

The clearest way to show this is to directly evaluate the scattering map $S$ on branch-point twist fields. The scattering map is a map acting on observables, ${\cal O}\mapsto S({\cal O})$, which represents the large-time forward evolution with $H$ and backward evolution with $H_0$ of local observables:
\eq{\label{eq:scattering}
S(\mathcal{O})=\lim_{t\rightarrow\infty}e^{-iH_0t}e^{iHt}\mathcal{O}e^{-iHt}e^{iH_0t}.
}
It is a simple matter to see that it allows to write the steady state using the original state:
\eq{
	\langle {\cal O}\rangle_{\text{ness}} = \langle
	S({\cal O}) \rangle_0.
}

The observables resulting from the application of the scattering map are to be evaluated in the state $\langle\cdot\rangle_0$. In this state, the left and right regions of the system are separated. The boundary conditions for both left and right regions, at the point $x=0$, are invariant under permutation of the replica copies. Hence, the unitary symmetry operator $Z$ for cyclic permutation, $\varphi_i(x)Z = Z\varphi_{i+1}(x)$, can be separated into two unitary operators generating the independent symmetries on the left and the right subsystems: $Z=Z_l\,Z_r$ with
\[
	\varphi_i(x) Z_{l} = \left\{\begin{array}{ll}
	Z_l\, \varphi_i(x) & (x>0) \\
	Z_l \,\varphi_{i+1}(x) & (x<0)
	\end{array}\right.,\quad
	\varphi_i(x) Z_{r} = \left\{\begin{array}{ll}
	Z_r \,\varphi_{i+1}(x) & (x>0) \\
	Z_r \,\varphi_{i}(x) & (x<0)
	\end{array}\right.,\quad
	Z_l^\dag Z_l = Z_r^\dag Z_r = {\bf 1}.
\]
\begin{subequations}\label{eq:twistfields-reversed}
For later convenience, we define the \textit{reversed twist field},
\eq{
\varphi_i(y,t)\,\mathcal{U}(x,t)=\left\{
\begin{array}{ll}
	\mathcal{U}(x,t)\,\varphi_{i}(y,t)&(x<y) \\ \mathcal{U}(x,t)\,\varphi_{i+1}(y,t)\qquad&(y<x)
\end{array}
\right.
}
with the corresponding reversed anti-twist field
\eq{
\varphi_i(y,t)\,\tilde{\mathcal{U}}(x,t)=\left\{
\begin{array}{ll}
	\tilde{\mathcal{U}}(x,t)\,\varphi_{i}(y,t)&(x<y) \\ \tilde{\mathcal{U}}(x,t)\,\varphi_{i-1}(y,t)\qquad&(y<x)
\end{array}
\right.
}
By definition, these are related to the normal twist fields via relations such as
$\mathcal{U}(x)=Z\atwist[](x)$, $\tilde{U}(x)=\tilde{Z}\twist[](x)$, etc. where $Z$ and $\tilde{Z}$ are the operators that permute the sheets one way or the other (i.e. they insert a branch cut over the entire length of the system).
\end{subequations}

From these relations, it is clear that
\begin{align}
\twist[](x_1)\atwist[](x_2)&=\tilde{\mathcal{U}}(x_1)\mathcal{U}(x_2)\\
\atwist[](x_1)\twist[](x_2)&=\mathcal{U}(x_1)\tilde{\mathcal{U}}(x_2).
\end{align}

Naturally, it is possible to identify $Z_{l}$ and $Z_{r}$ with appropriate regularizations of $\mathcal{U}(0^-)$ and $\twist[\;](0^+)$, respectively, with respect to the state $\langle\cdot\rangle_0$. For instance, $Z_{r} = \lim_{x\to0^+} x^{\Delta_n} \twist[\;](x)$.

We will obtain the following:
\begin{eqnarray}\label{stp}
S(\twist[]^+(x))&=&\left\{
\begin{array}{ll} \mathcal{U}^{-}(-x)&(x> 0)\\ Z_l\,\tilde{\mathcal{U}}^+(x)& (x<0) \end{array}\right. \\
S(\twist[]^-(x))&=&\left\{\begin{array}{ll}
	\twist[]^-(x)&(x\geq 0)\\ Z_r\atwist[]^+(-x) & (x<0)
\end{array}\right. \\
S(\mathcal{U}^{+}(x))&=&\left\{\begin{array}{ll}
	Z_l \,\tilde{\mathcal{U}}^{-}(-x) & (x>0)\\ \mathcal{U}^{+}(x) & (x\leq 0)
\end{array}\right.\\
S(\mathcal{U}^{-}(x))&=&\left\{\begin{array}{ll}
	Z_r \atwist[]^{-}(x) & (x>0)\\ \twist[]^+(-x) & (x\leq 0)
\end{array}\right.
\end{eqnarray}
as well as similar equations with the exchange $\twist[]^\pm,\mathcal{U}^\pm,Z_{l,r} \leftrightarrow \atwist[]^\pm,\tilde{\mathcal{U}}^{\pm},Z_{l,r}^\dag$.

One can interpret this map by analyzing its action on the cuts emanating from the positions of the twist fields. These cuts are to be divided into segments that fall into one of four regions: (A) left-moving fields in the left system, (B) left-moving fields in the right system, (C) right-moving fields in the left system, and (D) right-moving fields in the right system.
From the above equations, we note that, essentially, the map $S$ takes (B) and (C) into themselves, and (A) and (D) into each other.
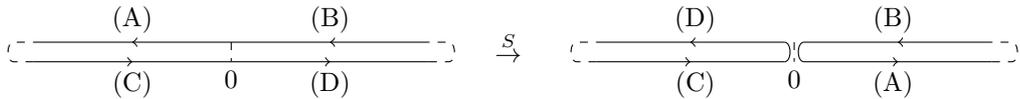
\begin{figure}[h!]
\centering\tikz[scale=1.3,anchor=base,baseline]{
\begin{scope}[decoration={
    markings,
    mark=at position 0.5 with {\arrow{>}}}]
\draw[thin,postaction={decorate}]    (2,0.2)--(1,0.2)node[anchor=south]{(B)}--(0,0.2);
\draw[thin,postaction={decorate}]    (0,0)node[anchor=north]{$0$}--(1,0)node[anchor=north]{(D)}--(2,0);
\draw[thin,postaction={decorate}]    (-2,0)--(-1,0)node[anchor=north]{(C)}--(0,0);
\draw[thin,postaction={decorate}]    (0,0.2)--(-1,0.2)node[anchor=south]{(A)}--(-2,0.2);
\draw[thin, dashed]
(0,0.2)--(0,0)
(-2,0)--(-2.2,0)to[in=180,out=180](-2.2,0.2)--(-2,0.2)
(2,0)--(2.2,0)to[in=0,out=0](2.2,0.2)--(2,0.2);
\end{scope}
}\hspace{0.5cm}$\stackrel{S}{\rightarrow}$\hspace{0.5cm}
\tikz[scale=1.3,anchor=base,baseline]{
\begin{scope}[decoration={
    markings,
    mark=at position 0.5 with {\arrow{>}}}]
\draw[thin,postaction={decorate}]    (2,0.2)--(1,0.2)node[anchor=south]{(B)}--(0.1,0.2);
\draw[thin,postaction={decorate}]    (0.1,0)--(1,0)node[anchor=north]{(A)}--(2,0);
\draw[thin,postaction={decorate}]    (-2,0)--(-1,0)node[anchor=north]{(C)}--(-0.1,0);
\draw[thin,postaction={decorate}]    (-0.1,0.2)--(-1,0.2)node[anchor=south]{(D)}--(-2,0.2);
\draw[thin]
(-0.1,0)to[in=0,out=0](-0.1,0.2)
(0.1,0)to[in=180,out=180](0.1,0.2);
\draw[thin, dashed]
(0,0.2)--(0,0)node[anchor=north]{$0$}
(-2,0)--(-2.2,0)to[in=180,out=180](-2.2,0.2)--(-2,0.2)
(2,0)--(2.2,0)to[in=0,out=0](2.2,0.2)--(2,0.2);
\end{scope}
}
\end{figure}

The map $S$ above immediately implies chiral factorization in the steady state: right- and left-movers are mapped onto left- and right-subsystems, respectively, which are independently thermalized in $\langle\cdot\rangle_0$. We may now map each independent subsystem onto a chiral theory, with in particular $Z_{l,r}$ mapping to the chiral replica permutation operator $Z$. One can see that the composition with $S$ is the identity operator, showing \eqref{chfa}.

It is then a simple matter to observe that the steady-state values of the entanglement entropy and negativity do reproduce the large-time limits evaluated by direct time evolution in the previous section.

\subsection{Calculation of the scattering map}

\begin{figure}[h!]
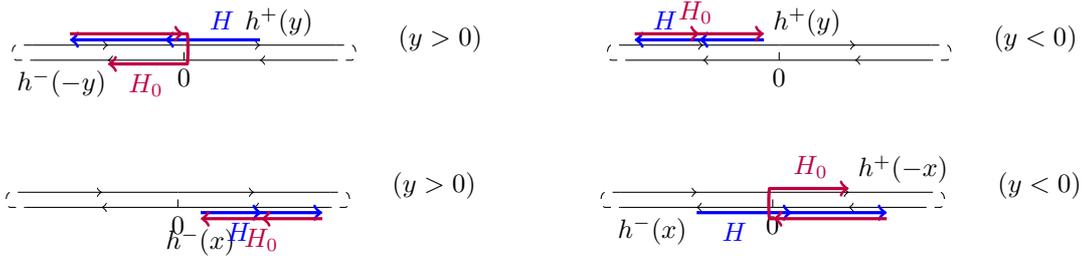

 \centering\tikz[scale=1,anchor=base,baseline]{
\begin{scope}[decoration={
    markings,
    mark=at position 0.5 with {\arrow{>}}}]
\draw[thin,postaction={decorate}]    (2,-0.2)--(0,-0.2);
\draw[thin,postaction={decorate}]    (0,0)--(1.25,0)node[anchor=south]{$h^{+}(y)$}--(2,0);
\draw[thin,postaction={decorate}]    (-2,0)--(0,0);
\draw[thin,postaction={decorate}]    (0,-0.2)node[anchor=north]{$0$}--(-0.9,-0.2)node[anchor=north east]{$h^{-}(-y)$}--(-2,-0.2);
\draw[thin, dashed]
(0,-0.2)--(0,0)
(-2,0)--(-2.2,0)to[in=180,out=180](-2.2,-0.2)--(-2,-0.2)
(2,0)--(2.2,0)to[in=0,out=0](2.2,-0.2)--(2,-0.2);
\draw[->,very thick,rounded corners,postaction={decorate},color=blue]
(1,0.07)--(0.5,0.07)node[anchor=south]{$H$}--(-1.5,0.07);
\draw[->,very thick,rounded corners,postaction={decorate},color=purple]
(-1.5,0.15)--(-0.1,0.15)to[in=0,out=0](-0.1,-0.25)--(-0.5,-0.25)node[anchor=north]{$H_0$}--(-1,-0.25);
\end{scope}
}
\hspace{0.3cm}
$(y>0)$\hspace{1.5cm}
\tikz[scale=1,anchor=base,baseline]{
\begin{scope}[decoration={
    markings,
    mark=at position 0.5 with {\arrow{>}}}]
\draw[thin,postaction={decorate}]    (2,-0.2)--(0,-0.2);
\draw[thin,postaction={decorate}]    (0,0)--(2,0);
\draw[thin,postaction={decorate}]    (-2,0)--(-0.2,0)node[anchor=south west]{$h^{+}(y)$}--(0,0);
\draw[thin,postaction={decorate}]    (0,-0.2)node[anchor=north]{$0$}--(-2,-0.2);
\draw[thin, dashed]
(0,-0.2)--(0,0)
(-2,0)--(-2.2,0)to[in=180,out=180](-2.2,-0.2)--(-2,-0.2)
(2,0)--(2.2,0)to[in=0,out=0](2.2,-0.2)--(2,-0.2);
\draw[->,very thick,rounded corners,postaction={decorate},color=blue]
(-0.2,0.07)--(-1.5,0.07)node[anchor=south]{$H$}--(-1.9,0.07);
\draw[->,very thick,rounded corners,postaction={decorate},color=purple]
(-1.9,0.15)--(-1.1,0.15)node[anchor=south]{$H_0$}--(-0.2,0.15);
\end{scope}
}
\hspace{0.3cm}
$(y<0)$\\
\vspace{0.5cm}
\tikz[scale=1,anchor=base,baseline]{
\begin{scope}[decoration={
    markings,
    mark=at position 0.5 with {\arrow{>}}}]
\draw[thin,postaction={decorate}]    (2,-0.2)--(0,-0.2);
\draw[thin,postaction={decorate}]    (0,0)--(2,0);
\draw[thin,postaction={decorate}]    (-2,0)--(-0.2,0)--(0,0);
\draw[thin,postaction={decorate}]    (0,-0.2)node[anchor=north]{$0$}--(-2,-0.2);
\fill
(0.3,-0.35)node[anchor=north]{$h^{-}(x)$};
\draw[thin, dashed]
(0,-0.2)--(0,0)
(-2,0)--(-2.2,0)to[in=180,out=180](-2.2,-0.2)--(-2,-0.2)
(2,0)--(2.2,0)to[in=0,out=0](2.2,-0.2)--(2,-0.2);
\draw[->,very thick,rounded corners,postaction={decorate},color=blue]
(0.3,-0.27)--(0.8,-0.27)node[anchor=north]{$H$}--(1.9,-0.27);
\draw[->,very thick,rounded corners,postaction={decorate},color=purple]
(1.9,-0.35)--(1.1,-0.35)node[anchor=north]{$H_0$}--(0.3,-0.35);
\end{scope}
}
\hspace{0.3cm}
$(y>0)$\hspace{1.5cm}
\tikz[scale=1,anchor=base,baseline]{
\begin{scope}[decoration={
    markings,
    mark=at position 0.5 with {\arrow{>}}}]
\draw[thin,postaction={decorate}]    (2,-0.2)--(0,-0.2);
\draw[thin,postaction={decorate}]    (0,0)--(1,0)node[anchor=south west]{$h^{+}(-x)$}--(2,0);
\draw[thin,postaction={decorate}]    (-2,0)--(0,0);
\draw[thin,postaction={decorate}]    (0,-0.2)node[anchor=north]{$0$}--(-1,-0.2)node[anchor=north east]{$h^{-}(x)$}--(-2,-0.2);
\draw[thin, dashed]
(0,-0.2)--(0,0)
(-2,0)--(-2.2,0)to[in=180,out=180](-2.2,-0.2)--(-2,-0.2)
(2,0)--(2.2,0)to[in=0,out=0](2.2,-0.2)--(2,-0.2);
\draw[->,very thick,rounded corners,postaction={decorate},color=blue]
(-1,-0.27)--(-0.5,-0.27)node[anchor=north]{$H$}--(1.5,-0.27);
\draw[->,very thick,rounded corners,postaction={decorate},color=purple]
(1.5,-0.35)--(0.1,-0.35)to[in=180,out=180](0.1,0.05)--(0.5,0.05)node[anchor=south]{$H_0$}--(1,0.05);
\end{scope}
}
\hspace{0.3cm}
$(y<0)$\
\caption{
The forward time evolution with $H$ and subsequent backward time evolution with $H_0$ of $h^\text{+}(y)$ in equation \eqref{eq:scattering_r} and $h^\text{-}(x)$ in equation \eqref{eq:scattering_l} is shown.
}
\label{fig:scattering}
\end{figure}

We deduce the form of $S(\twist[])$ and $S(\atwist[\;])$ by comparing the equal-time exchange relations before and after the process of forward- and backward time evolution. Consider the right- and left-moving energy densities $h^\pm(y)$. On these fields, the scattering map is given by \cite{}
\eq{\label{eq:scattering_r}
S(h^{+}(y))=\left\{
\begin{array}{ll}
	h^{-}(-y)\qquad &(y>0)\\ h^{+}(y)&(y<0)
\end{array}\right.
}
and
\eq{\label{eq:scattering_l}
S(h^{-}(y))=\left\{
\begin{array}{ll}
	h^{-}(y)&(y>0)\\ h^{+}(-x)\qquad &(y<0)
\end{array}\right.
}
See Fig. \ref{fig:scattering} for a depiction of these results.

Consider the equal-time exchange relations
\eq{
h^{+}_{i}(y)\twist[+](0)=\left\{
\begin{array}{ll}
	\twist[+](0)\,h^{+}_{i}(y)&(y<x) \\ \twist[+](0)\,h^{+}_{i+1}(y).\qquad&(y>x)
\end{array}
\right.
}
Since the scattering \eqref{eq:scattering} is a conjugation, the exchange relations are preserved. Therefore, we also have:
\eq{
S(h^{+}_{i}(y))S(\twist[+](x))=\left\{
\begin{array}{ll}
	S(\twist[+](x))\,S(h^{+}_{i}(y))&(y<x) \\ S(\twist[+](x))\,S(h^{+}_{i+1}(y))\qquad&(y>x)
\end{array}
\right.
}
which, using \eqref{eq:scattering_r}, gives the following equations:
\begin{alignat}{3}
h^{+}_{i}(y)S(\twist[+](x))&=
	S(\twist[+](x))\,h^{+}_{i}(y)\qquad &(y<x\;\&\;y<0)\\
h^{+}_{i}(y)S(\twist[+](x))&=
	S(\twist[+](x))\,h^{+}_{i+1}(y)\qquad &(y>x\;\&\;y<0)\\
h^{-}_{i}(-y)S(\twist[+](x))&=
	S(\twist[+](x))\,h^{-}_{i}(-y)\qquad\qquad &(y<x\;\&\;y>0)\\
h^{-}_{i}(-y)S(\twist[+](x))&=
	S(\twist[+](x))\,h^{-}_{i+1}(-y)\qquad\qquad &(y>x\;\&\;y>0)
\end{alignat}
We can find more information on $S(\twist[+])$ by doing the same with the exchange relations with $h^{-}$. Since $\twist[+]$ only acts as boundary-changing field on right-moving fields, these exchange relations are trivial:
\eq{
h^{-}_i(y)\twist[+](x)=\twist[+](x)h^{-}_i(y).
}
After scattering, these are
\begin{alignat}{3}
h^{-}_i(y)S(\twist[+](x))&=S(\twist[+](x))h^{-}_i(y) \qquad\qquad& (y>0)\\
h^{+}_i(-y)S(\twist[+](x))&=S(\twist[+](x))h^{+}_i(-y)& (y<0)
\end{alignat}
Collecting these equations, we have the following exchange relations for $S(\twist[+])$:
\begin{align}
h^{+}_i(y)S(\twist[+](x))&=\left\{
\begin{array}{ll}
S(\twist[+](x))h^{+}_i(y)\qquad\;\;\,&(y<x\;\&\;y<0,\; y>0)\\ S(\twist[+](0))h^{+}_{i+1}(y)&(x<y<0)
\end{array}
\right.\\
h^{-}_i(y)S(\twist[+](x))&=\left\{
\begin{array}{ll}
S(\twist[+](0))h^{-}_i(y)&(-x<y<0,\;y>0)\\ S(\twist[+](0))h^{-}_{i+1}(y)\qquad&(y<-x\;\&\;y<0)
\end{array}
\right.
\end{align}
When $x\geq 0$, this just simplifies to:
\begin{align}
h^{+}_i(y)S(\twist[+](x))&=
	S(\twist[+](x))h^{+}_i(y)\\
h^{-}_i(y)S(\twist[+](x))&=
	\left\{\begin{array}{ll}S(\twist[+](x))h^{-}_i(y) & (-x<y)\\
S(\twist[+](x))h^{-}_{i+1}(y)\qquad & (y<-x)\end{array}\right.
\end{align}
which means that:
\eq{
S(\twist[+](x))=\mathcal{U}_{-}(-x)\qquad x\geq 0
}
However, for $x<0$, we have
\begin{align}
h^{+}_i(y)S(\twist[+](x))&=
	\left\{\begin{array}{ll}S(\twist[+](x))h^{+}_i(y) & (y<x)\\ S(\twist[+](x))h^{+}_{i+1}(y)\qquad & (x< y<0)\\ S(\twist[+](x))h^{+}_i(y) & (y>0)\end{array}\right.\\
h^{-}_i(y)S(\twist[+](x))&=
	\left\{\begin{array}{ll}
S(\twist[+](x))h^{-}_{i+1}(y)\qquad & (y<0)\\
S(\twist[+](x))h^{-}_i(y) & (y>0)\end{array}\right.
\end{align}
so for $x<0$ the result of $S(\twist[+])$ is a product between $\mathcal{U}_{-}(0)$ and a branch cut between $x<y<0$ for the right-moving fields:
\eq{
S(\twist[+](x))\propto\mathcal{U}_{-}(0)\twist[+](x)\atwist[+](0)\qquad (x<0)
}
This can be re-written as
\eq{
S(\twist[+](x))\propto\mathcal{U}_{-}(0)\tilde{\mathcal{U}}^+(x) \mathcal{U}^+(0) \propto Z_l \,\tilde{\mathcal{U}}^+(x) \qquad (x<0)
}
Taking into account the normalization of the field, we obtain \eqref{stp}.

The other relations can be obtained similarly.

\section{Relations between structure constants}
\label{sec:structureconstants}

In this appendix we will establish general relations amongst structure constants associated with cyclic permutation twist fields (powers of the branch-point twist fields). One of these relations will be \eqref{eqCC}, needed in the main text. We will use the following notation for the fields that permute the $n$ sheets cyclically by an amount of $i\in
\{0,1,\ldots,n-1\}$:
\eq{
\sigma_i:=\twist^i.
}
For convenience, we extend the notation to $i\in\mathbb{N}$ by periodicity $\sigma_{i+n} = \sigma_i$.

By the fundamental properties of twist fields, the OPE $\sigma_i(x)\sigma_j(y)$ must be in the twist sector associated to $i+j$, hence
\eq{\label{OPEij}
\sigma_i(x)\sigma_j(y)\stackrel{x\rightarrow y}{\sim}\frac{C_{\sigma_i,\sigma_j}^{\sigma_{i+j}}\,\sigma_{i+j}(y)}{|x-y|^{d_i+d_j-d_{i+j}}}
}
where $d_i$ is the scaling dimension of $\sigma_i$. We wish to establish constraints on the structure constants $C_{i,j} := C_{\sigma_i,\sigma_j}^{\sigma_{i+j}}$.

The CFT normalization is the condition
\eq{
C_{i,-i}=1
}
which amounts to normalizing the two-point function as
\eq{\label{eq:twopoint}
\langle\sigma_i(x_1)\sigma_j(x_2)\rangle=\delta_{i+j,0}\frac{1}{x_{12}^{2d_i}}.
}
Further, since the subgroup of cyclic permutation is abelian, we have
\eq{
	C_{i,j} = C_{j,i}.
}
By acting with the permutation element that inverts all sheets (this elements generates, along with cyclic permutation elements, the maximal subgroup that preserves the cyclic permutation subgroup), we also find
\eq{
	C_{-i,-j} = C_{i,j}.
}
With the identity element, the OPE is trivial, whence
\eq{
	C_{0,i} = 1.
}

In order to obtain additional information, we consider three-point functions, which are fixed by conformal invariance up to multiplicative constants:
\eq{\label{eq:threepoint}
\langle\sigma_i(x_1)\sigma_j(x_2)\sigma_k(x_3)\rangle=\delta_{i+j+k,0}\frac{C_{\sigma_i,\sigma_j,\sigma_{-i-j}}}{x_{12}^{d_i+d_j-d_k}x_{13}^{d_i+d_k-d_j}x_{23}^{d_j+d_k-d_i}}
}
From the OPE \eqref{OPEij}, we find
\eq{
\langle\sigma_i(x_1)\sigma_j(x_2)\sigma_k(x_3)\rangle\stackrel{x_1\rightarrow x_2}{\sim}\frac{C_{i,j}\langle\sigma_{i+j}(x_2)\sigma_k(x_3)\rangle}{x_{12}^{d_i+d_j-d_{i+j}}}
=\delta_{i+j+k,0}\frac{C_{i,j}}{x_{12}^{d_j+d_j-d_{i+j}}x_{23}^{d_{i+j}+d_k}}.
}
and comparing with the limit $x_1\rightarrow x_2$ in \eqref{eq:threepoint}, it is clear that $C_{\sigma_i,\sigma_j,\sigma_{-i-j}}=C_{i,j}$. Similarly,
\eq{
\langle\sigma_i(x_1)\sigma_j(x_2)\sigma_k(x_3)\rangle\stackrel{x_2\rightarrow x_3}{\sim}\frac{C_{j,k}\langle\sigma_{i}(x_1)\sigma_{j+k}(x_3)\rangle}{x_{23}^{d_j+d_k-d_{j+k}}}
=\delta_{i+j+k,0}\frac{C_{j,k}}{x_{23}^{d_j+d_k-d_{j+k}}x_{13}^{d_i+d_{j+k}}},
}
wherefore $C_{\sigma_i,\sigma_j,\sigma_{-i-j}}=C_{j,-i-j}$. From these we obtain an extra constraint on the structure constants:
\eq{\label{eq:strcst}
C_{i,j}=C_{j,-i-j}.
}
Putting $i=j=1$ in this equation, we find \eqref{eqCC}.

Finally, by factorization of the multi-sheeted theory, we also have
\eq{
C_{i/k,j/k;n/k}=C_{i,j;n}
}
where we have explicitly written the number $n$ of copies via $C_{i,j} = C_{i,j;n}$, and where $k$ divides $i$, $j$ and $n$.

Using these relations, one can reduce the number of unkonwn structure constants. For instance, for $n=2$ we have $C_{i,j}=1$ for all $i,j$. For $n=3$, we find a single unknown structure constant, $C_{1,1}=C_{2,2}$ (the others are unity). For $n=4$, there is also a single unknown structure constant, $C_{1,1}=C_{1,2}=C_{2,1}=C_{2,3}=C_{3,2}=C_{3,3}$ (the others are unity). The non-unity structure constants for $n=5$ are $C_{1,1}=C_{1,3}=C_{2,4}=C_{4,4}$ and $C_{1,2}=C_{2,2}=C_{3,3}=C_{3,4}$ (up to equality under exchanging indices). For $n=6$, the unknown are $C_{1,1}=C_{1,4}=C_{2,5}=C_{5,5}$ and $C_{1,2}=C_{1,3}=C_{2,3}=C_{3,4}=C_{3,5}=C_{4,5}$ (again up to equality under exchanging indices), $C_{2,2}=C_{4,4}$ are related to structure constants for $n=3$, and the others are unity. The unknown structure constants must in general be determined by evaluating 4-point functions, and depend on the particular CFT model under consideration.

\section{Boundary entropy}
\label{sec:boundary_entropy}
In order to find an expression for the boundary entropy introduced in \cite{AffleckLudwig1991} in terms of the nonuniversal constants appearing in our expressions for the EE and the logarithmic negativity, we must relate the EE for an interval of length $\ell$ on the half line to the EE for an interval of length $2\ell$ on the line, both in equilibrium. Not specifying the state of the system, we first find an expression for the trace of the system with a boundary:
\eq{
\text{Tr}\rho^{\text{bdry}}_{[0,\ell]}=c_n\delta^{2d_n}\lim_{\varepsilon\rightarrow 0}\langle\twist(\varepsilon)\atwist(\ell)\rangle_{\mathbb{R}^+}.
}
We can compute this by unfolding \eqref{eq:unfoldingmap} and using the OPE $\chatwist(-\varepsilon)\chtwist(\varepsilon)\sim (2\varepsilon)^{-2\cdim}$,
\eq{
\lim_{\varepsilon\rightarrow 0}\langle\twist(\varepsilon)\atwist(\ell)\rangle_{\mathbb{R}^+}=\lim_{\varepsilon\rightarrow 0}\langle\chtwist(-\ell)\chatwist(-\varepsilon)\chtwist(\varepsilon)\chatwist(\ell)\rangle^{ch}_{\mathbb{R}}=\lim_{\varepsilon\rightarrow 0}(2\varepsilon)^{-2\cdim}\langle\chtwist(-\ell)\chatwist(\ell)\rangle^{ch}_{\mathbb{R}}.
}
As discussed in section~\ref{sec:twistfields}, in taking the limit $\varepsilon\rightarrow 0$ we are in effect exchanging this limit with the scaling limit. In order to account for this, we define a constant $b_n:=2\varepsilon/\delta$, in terms of which the trace can be written as
\eq{
\text{Tr}\rho^{\text{bdry}}_{[0,\ell]}=c_nb_n^{-2\cdim}\delta^{2\cdim}\langle\chtwist(-\ell)\chatwist(\ell)\rangle^{ch}=c_n^{1/2}b_n^{-d_n}\left(\text{Tr}\rho^{\text{bulk}}_{[-\ell,\ell]}\right)^{1/2}.
}
This leads to the relation
\eq{
S^{\text{bdry}}_{[0,\ell]}=\frac{1}{2}S^{\text{bulk}}_{[-\ell,\ell]}+\frac{c'_1}{2}-\frac{c}{12}\ln b_1,
}
from which we conclude that the boundary entropy is equal to \cite{CalabreseCardy2004,LaflorencieSorensenChangAffleck2006,ZhouBarthelFjaerestadSchollwoeck2006},
\eq{\label{eq:affleckentropy}
\ln g=-\frac{c}{12}\ln b_1.
}
Here we note that whenever we encounter $b_1$, we take this to be the limit $\lim_{n\rightarrow 1}b_n$. Recall that $b_n$, defined in \eqref{eq:b_n}, appears when taking a limit where a bulk field goes to the boundary, which amounts to exchanging limits $\delta\rightarrow 0$ and $\varepsilon\rightarrow 0$. For $n=1$ this works for any $b_1$, since the branch-point twist fields are just the identity, and do not depend on position.

\section{Mutual information}
\label{sec:mutualinfo}

The mutual information between two regions $A_1$ and $A_2$ is defined as
\eq{
I_{A_1,A_2}:=S_{A_1}+S_{A_2}-S_{A_1\cup A_2},
}
with $S_A$ the EE between a region $A$ and the rest of the system, and similaryl the R\'{e}nyi mutual information is defined by
\eq{
I^{(n)}_{A_1,A_2}:= S^{(n)}_{A_1}+S^{(n)}_{A_2}-S^{(n)}_{A_1\cup A_2}.
}
In order to find relations for the mutual information after a local quench, we first must compute relations similar to \eqref{eq:renyi_u<t<v} and \eqref{eq:renyi_t>v} but in the case of an interval $[-v,-u]$ in the left system, and the case of an interval $[-v,-u]\cup[u,v]$ in both systems. We take $u>0$ and $v>0$, as we did before. The relation for $S^{(n)}_{[-v,-u]}$ can be obtained by simply replacing $\beta_l\leftrightarrow\beta_r$ in \eqref{eq:renyi_u<t<v} and \eqref{eq:renyi_t>v}. In order to find similar relations for $S^{(n)}_{[-v,-u]\cup[u,v]}$, we follow the method outlined in section~\ref{sec:twistfields}. First, we note that $t>v$ the cuts do not extend across the defects, wherefore the NESS result should be time independent. For times $u<t<v$, the cuts do extend across the defects, and we have to regularize the expression for the trace by inserting extra pairs of twist fields as follows: 
\begin{multline}
\text{Tr}\rho_{[-v,-u]\cup[u,v]}^n(u<t<v)\\
=c_n^2\delta^{4d_n}\lim_{\varepsilon\rightarrow 0}(2\varepsilon)^{4d_n}\langle\twist(-v,t)\atwist(-t-\varepsilon,t)\twist(-t+\varepsilon,t)\atwist(-u,t)\twist(u,t)\atwist(t-\varepsilon,t)\twist(t+\varepsilon,t)\atwist(v,t)\rangle_0.
\end{multline}
Performing the time evolution on the chiral twist fields defined in section~\ref{sec:twistfields} and using the necessary OPEs, we obtain
\begin{multline}
\text{Tr}\rho_{[-v,-u]\cup[u,v]}^n(u<t<v)=c_n^2\delta^{8\cdim}\lim_{\varepsilon\rightarrow 0}(2\varepsilon)^{4\cdim}\\
\langle\twist^{+}(-v-t,0)\twist^{-}(-v+t,0)
\atwist^{+}(-u-t,0)\twist^{+}(u-t,0)\atwist^{-}(-\varepsilon,0)\atwist^{+}(-\varepsilon,0)\rangle_{\lt}\\
\langle\twist^{-}(\varepsilon,0)\atwist^{-}(-u+t,0)\twist^{-}(u+t,0)
\twist^{+}(\varepsilon,0)\atwist^{+}(v-t,0)\atwist^{-}(v+t,0)\rangle_{\rt}.
\end{multline}
After the unfolding map, this becomes
\begin{multline}
\text{Tr}\rho_{[-v,-u]\cup[u,v]}^n(u<t<v)=c_n^2\delta^{8\cdim}\lim_{\varepsilon\rightarrow 0}(2\varepsilon)^{4\cdim}
\langle\chtwist(-v-t)
\chatwist(-u-t)\chtwist(u-t)\chatwist(-\varepsilon)\chtwist(\varepsilon)\chatwist(v-t)\rangle^{ch}_{\beta_l}\\
\langle\chtwist(-v-t)\chatwist(-u-t)\chtwist(u-t)\chatwist(-\varepsilon)\chtwist(\varepsilon)\chatwist(v-t)\rangle^{ch}_{\beta_r}.
\end{multline}
We can use the OPEs again to simplify this equation further:
\begin{multline}
\text{Tr}\rho_{[-v,-u]\cup[u,v]}^n(u<t<v)=c_n^2\delta^{8\cdim}
\langle\chtwist(-v-t)
\chatwist(-u-t)\chtwist(u-t)\chatwist(v-t)\rangle^{ch}_{\beta_l}\\
\langle\chtwist(-v-t)\chatwist(-u-t)\chtwist(u-t)\chatwist(v-t)\rangle^{ch}_{\beta_r},
\end{multline}
leading to the relation
\eq{
\text{Tr}\rho_{[-v,-u]\cup[u,v]}^n(u<t<v)=\left(\text{Tr}(\rho^{eq}_{[-v,-u]\cup[u,v]})^n(\beta_l)\right)^{1/2}\left(\text{Tr}(\rho^{eq}_{[-v,-u]\cup[u,v]})^n(\beta_r)\right)^{1/2}.
}
Note that this relation does not depend on time. Since we also know that for times $t<u$ and $t>u$ the relation is time independent, we find that the relation between the R\'{e}nyi entropy after the local quench and equilibrium quantities for two intervals of equal length, at equal distances from the point of connection, is given by
\eq{\label{eq:renyirelation}
S^{(n)}_{[-v,-u]\cup[u,v]}(t;\beta_l,\beta_r)=\frac{1}{2}\left(S^{(n),eq}_{[-v,-u]\cup[u,v]}(\beta_l)+S^{(n),eq}_{[-v,-u]\cup[u,v]}(\beta_r)\right),
}
where, as before, we explicitly label equilibrium expressions. So the EE after a local quench between $[-v,-u]\cup[u,v]$ and its complement, does not depend on time.
Using \eqref{eq:renyirelation} and \eqref{eq:renyi_u<t<v} (both as is and with $\beta_l\leftrightarrow\beta_r$), we have for $u<t<v$ the following relation for the R\'{e}nyi mutual information:
\eq{
I^{(n)}_{[-v,-u],[u,v]}(u<t<v;\beta_l,\beta_r)=\frac{1}{2}\left(I^{(n),eq}_{[u,t],[-v,-u]\cup[t,v]}(\beta_l)+I^{(n),eq}_{[u,t],[-v,-u]\cup[t,v]}(\beta_r)\right)-c'_n+\frac{2d_n}{1-n}\ln b_n,
}
with $c'_n:=\ln c_n/(1-n)$. The mutual information for this time regime simplifies to
\eq{
I_{[-v,-u],[u,v]}(u<t<v;\beta_l,\beta_r)=\frac{1}{2}\left(I^{eq}_{[u,t],[-v,-u]\cup[t,v]}(\beta_l)+I^{eq}_{[u,t],[-v,-u]\cup[t,v]}(\beta_r)\right)+c'_1+2\ln g,
}
with $\ln g$ the boundary entropy first discussed in \cite{AffleckLudwig1991}, see Appendix~\ref{sec:boundary_entropy}
In the NESS, we can use \eqref{eq:renyirelation} together with \eqref{eq:renyi_t>v} (again, for both choices of $\beta_l,\beta_r$) to find the simpler relation:
\eq{
I^{(n)}_{[-v,-u],[u,v]}(t>v;\beta_l,\beta_r)=\frac{1}{2}\left(I^{(n),eq}_{[-v,-u],[u,v]}(\beta_l)+I^{(n),eq}_{[-v,-u],[u,v]}(\beta_r)\right),
}
with the mutual information given by
\eq{
I_{[-v,-u],[u,v]}(t>v;\beta_l,\beta_r)=\frac{1}{2}\left(I^{eq}_{[-v,-u],[u,v]}(\beta_l)+I^{eq}_{[-v,-u],[u,v]}(\beta_r)\right),
}

\bibliography{Entanglement}{}
\bibliographystyle{ieeetr}


\end{document}